\documentclass[twocolumn]{aastex63}

\usepackage{amsfonts,amsmath,graphicx,natbib,subfigure,color,gensymb,longtable,footnote,soul}

\received{}
\revised{ -- }
\accepted{}
\submitjournal{ApJ}

\shorttitle{Star formation and morphological properties of P\lowercase{an}-STARRS galaxies}
\shortauthors{Baldeschi et al. }

\graphicspath{{./}{figures/}}

\begin{document}

\title{Star formation and morphological properties of galaxies in the P\lowercase{an}-STARRS 3$\pi$ survey- I.\\ A machine learning approach to galaxy  and supernova classification}

\correspondingauthor{Adriano Baldeschi}
\email{adriano.baldeschi@northwestern.edu}

%%%%%%%%%%%

\author{Baldeschi A.}
\affiliation{Center for Interdisciplinary Exploration and Research in Astrophysics (CIERA) and Department of Physics and Astronomy, Northwestern University, Evanston, IL 60208}

\author{Miller A.}
\affiliation{Center for Interdisciplinary Exploration and Research in Astrophysics (CIERA) and Department of Physics and Astronomy, Northwestern University, Evanston, IL 60208}
\affiliation{The Adler Planetarium, Chicago, IL 60605, USA}

\author{Stroh M.}
\affiliation{Center for Interdisciplinary Exploration and Research in Astrophysics (CIERA) and Department of Physics and Astronomy, Northwestern University, Evanston, IL 60208}

\author[0000-0003-4768-7586]{Margutti R.}
\altaffiliation{Alfred P. Sloan Fellow.}
\affiliation{Center for Interdisciplinary Exploration and Research in Astrophysics (CIERA) and Department of Physics and Astronomy, Northwestern University, Evanston, IL 60208}
\affiliation{CIFAR Azrieli Global Scholars program, CIFAR, Toronto, Canada}

\author[0000-0001-5126-6237]{Coppejans D.L.}
\affiliation{Center for Interdisciplinary Exploration and Research in Astrophysics (CIERA) and Department of Physics and Astronomy, Northwestern University, Evanston, IL 60208}

%%%%%%%%%%%%%%%%%



\begin{abstract}
We present a classification of galaxies in the P\lowercase{an}-STARRS1 (PS1) $3\pi $ survey based on their recent star formation history and  morphology. Specifically, we train and test two  Random Forest (RF) classifiers using photometric features (colors and moments) from the PS1 data release 2. Labels for the morphological classification are taken from \cite{Huertas-Company+2011}, while labels for the star formation fraction (SFF) are from the \citet{Blanton2005} catalog. We find that colors provide more predictive accuracy than photometric moments. We morphologically classify galaxies as either early- or late-type, and our RF model achieves a 78\% classification accuracy. Our second model classifies galaxies as having either a low-to-moderate or high SFF. This model achieves an 89\% classification accuracy. We apply both RF classifiers to the entire PS1 $3\pi$ dataset, allowing us to assign two scores to each PS1 source: $P_\mathrm{HSFF}$, which quantifies the probability of having a high SFF, and $P_\mathrm{spiral}$, which quantifies the probability of having a late-type morphology. Finally, as a proof of concept, we apply our classification framework to supernova (SN) host-galaxies from the Zwicky Transient Factory and the Lick Observatory Supernova Search samples. We show that by selecting on $P_\mathrm{HSFF}$ or $P_\mathrm{spiral}$ it is possible to significantly enhance or suppress the fraction of core-collapse SNe (or thermonuclear SNe) in the sample with respect to random guessing. This result demonstrates how contextual information can aid transient classifications at the time of first detection. In the current era of spectroscopically-starved time-domain astronomy, prompt automated classification is paramount.
\end{abstract}


%%%%%%%%%%%%%%%%%%%%%%%%%%%%%%%%%%%%%%%%%%%
\section{Introduction} \label{sec:intro}
 
The improved sensitivity, cadence, and field of view of recent and current astronomical transient surveys  
have led to the discovery of new types of rare transients that sample the extremes of the luminosity and time-scale parameter space of cosmic explosions, and dramatically enhanced our understanding of classes of transients that were already known to exist. Classes of exotic transients that were recently discovered include Superluminous supernovae (SLSNe, e.g., 
\citealt{Quimby2011,Gal-Yam2019})  and Fast and Blue Optical Transients (FBOTs, e.g. \citealt{Drout2014,Arcavi2016, Tanaka2016, Shivvers2016,Pursiainen2018,Tampo2020}).
Furthermore, astronomical surveys are now capable of routinely discovering supernovae (SNe) within  one day (or less) of the explosion \citep[e.g.,][]{Gal-Yam2011}. Future surveys like the Legacy Survey of Space and Time (LSST, \citealt{Ivezic2019}) carried out on the  Vera C. Rubin Observatory will drastically increase  the discovery rate of new transients,  
which will make prompt spectroscopic classification of a sizeable fraction of transients effectively unfeasible.
Indeed,  even intrinsically rare events like SLSNe are expected to be detected at a rate of $\sim10^4\,\rm{yr^{-1}}$ \citep{Villar2018}. 
It is thus imperative to develop new pathways for prompt transient classification.

The  most common classification methods consist of leveraging the transient photometry by using state-of-the-art machine learning algorithms. A first generation of photometric transient classifiers (mostly developed in response to the Supernova Photometric Classification Challenge, SPCC, \citealt{Kessler2010,Kessler10b}) can be broadly divided into empirical template-fitting methods (e.g., \citealt{Sullivan2006,Sako2008,Sako2011}), and algorithms that rely on the derivation of (computationally expensive) features from extended photometry (e.g., \citealt{Newling2011,Karpenka2013,Moller2016,Lochner+2016,Sooknunan2018,Narayan2018,Ishida2019,Villar+2019}). 
A new generation of transient classifiers that do not require complete light-curve phase coverage and/or feature extraction have emerged in the last few years and have been applied to transient images (\citealt{Carrasco-Davis2019}) and SN photometric time series (\citealt{Charnock2017,Moss+2018,Pasquet2019,Muthukrishna+2019,Moller2020}). The advanced non-feature based neural network architectures used in these works include recurrent neural networks (RNN), and deep neural networks (DNN). 
More recently, \citet{Muthukrishna+2019} developed  \texttt{RAPID} (Real-time Automated Photometric IDentification) and tested this DNN-based tool on the PLAsTiCC data set. Differently to the other classification schemes above, \texttt{RAPID} employs a deep learning model that is able to promptly classify different types of transients with very limited light-curve information, without requiring complete phase coverage.\footnote{See \citep{Sullivan2006}  for earlier attempts to prompt transient classification with minimal light-curve information.}

Complementary methods for prompt transient classification are those that employ contextual information, i.e. the properties of the galactic environments where transients are discovered (e.g., \citealt{Foley+2013} and the \texttt{sherlock} package\footnote{https://github.com/thespacedoctor/sherlock}). These methods rely on very well known correlations between transient types and their host-galaxy environments (for example, core-collapse SNe tend to trace star formation, while type Ia SNe occur both in early-type and late-type host galaxies). 

Building on the early results from \citet{Foley+2013} that showed that, among other contextual properties,  the host-galaxy morphology has the largest predictive power for SN typing, we present the classification of all the galaxies detected by the Panoramic Survey Telescope and Rapid Response System (P\lowercase{an}-STARRS) $3\pi$ survey.  We adopt a random forest (RF) machine learning  approach that leverages galaxy features to classify the galaxies based on their morphology (i.e. elliptical/S0 vs.\ spiral) and their star formation properties.
We train and test two supervised machine learning RF algorithms starting from the morphological  classifications of SDSS galaxies by \cite{Huertas-Company+2011}, and the star formation properties from  \cite{Blanton2005,Blanton2007}. 
We present a catalog where for each P\lowercase{an}-STARRS source we provide two scores that quantify the probability of having spiral morphology ($P_\mathrm{spiral}$) and the probability of having a high recent star formation rate ($P_\mathrm{HSFF}$).
Finally, as a proof of concept we use both scores  to classify host galaxies of SNe from  the  Zwicky Transient Facility (ZTF) Bright Transient Survey (BTS, \citealt{Fremling2019}) and from the Lick Observatory Supernova Search (LOSS, \citealt{Leaman2011}) and we show how $P_\mathrm{spiral}$ and $P_\mathrm{HSFF}$ correlate with the fraction of core-collapse SNe (or thermonuclear SNe) in the sample. Our catalog of Pan-STARRS galaxies classifications will be publicly available and can be ingested by transient brokers (e.g., ALeRCE,\footnote{\cite{Smith19},  http://alerce.science}  ANTARES\footnote{\cite{Saha2014,Saha2016}, https://antares.noao.edu/}, LASAIR,\footnote{https://lasair.roe.ac.uk}) that sort, cross-reference, and value-add streams of alerts from astronomical surveys. The inferred host-galaxy properties can be used by recommender engines like \texttt{REFIT} \citep{Sravan2020}, together with the photometric information, to promptly inform decisions on transient follow-up  at the time of their first detection.

This paper is organized as follows. In \S\ref{datdesc} we describe the data sets used, while in \S\ref{preprocess}   we train and test the first RF model by cross-matching  the Pan-STARRS DR2 catalog with the \cite{Huertas-Company+2011} data set. In \S\ref{RFNYU} we train and test the second  RF model by cross-matching the P\lowercase{an}-STARRS DR2 catalog with the New York University Value-Added Galaxy Catalog (NYU-VAGC of \citealt{Blanton2005,Blanton2007}). In \S\ref{panst} we apply the second RF to the entire P\lowercase{an}-STARRS dataset. Finally, in  \S\ref{transient} we classify the host galaxies of ZTF and LOSS SNe by leveraging the results of \S\ref{panst}.
 Conclusions are drawn in \S\ref{conclusions}.

%%%%%%%%%%%%%%%%%%%%%%%%%%%%%%%%%%%%%%%%%%%%%%%
\section{Data sets description}
\label{datdesc}
For our analysis we use three data catalogs. Specifically, we  utilize  the second P\lowercase{an}-STARRS1  (PS1, \citealt{Chambers+2016}) data release  of the $3\pi $ survey (PS1-DR2 hereafter), the ``Huertas-Company data-set'' (HC hereafter,  \citealt{Huertas-Company+2011}), and the New York University Value-Added Galaxy Catalog (NYU-VAGC hereafter,   \citealt{Blanton2005,Blanton2007}). We used the HC, NYU-VAGC and  a subsample of PS1-DR2  for training and testing two RF algorithms that we later apply to the entire PS1-DR2 catalog. 

PS1 is a system for wide-field astronomical imaging  developed at the University of Hawaii and located on the island of Maui. PS1 data have been acquired with a 1.8 meter telescope and a 1.4 Gigapixel camera to capture images of the sky through five optical  filters ($g_{P1}$ [4866 {\AA}], $r_{P1}$ [6215 {\AA}], $i_{P1}$ [7545 {\AA}], $z_{P1}$ [8679 {\AA}], $y_{P1}$ [9633 {\AA}]). Two surveys have been carried out with the PS1 telescope: the medium deep survey and the 3$\pi$ survey (3$\pi$S). In this paper, we use data from the 3$\pi$S, which covers a larger fraction of the sky. The 3$\pi$S  covers the sky north of declination $\delta=  - 30^\circ$  with the five filters listed above, and includes  data acquired between 2009-06-02 and 2014-03-31. The maximum depth of the 3$\pi$S for the stack images is $\sim$23.5 mag for the $g_{P1}$, $r_{P1}$, and  $i_{P1}$ filters, while it is $\sim$22.5 and $\sim$21.5 mag for the $z_{P1}$ and $y_{P1}$ filters, respectively. 
There have been two data releases for the 3$\pi$S \citep{Chambers+2016}. 
Data Releases 1 and 2 (DR1 and DR2) include stacked images and a database with the photometry of all the sources detected in 3$\pi$S (both extended and point-like). DR2 additionally contains forced photometry for each epoch.
In this paper, we use the data from the  ``StackObjectAttributes'' table within DR2\footnote{\href{https://outerspace.stsci.edu/display/PANSTARRS/}{StackObjectAttributes table link}}.
This table
contains the photometric information (e.g., PSF-flux, Kron-flux) of the stacked  data, calculated as described in \cite{Magnier2013}.  To be included in this table, a source must be detected with a signal-to-noise $S/N>20$ in an individual exposure. There are often multiple detections of the same source from subsequent exposures, which means that there can be multiple photometric entries for a single source.  In \S\ref{ffs} we describe the columns of the StackObjectAttributes table that we used as features to train our RF classifiers.

 The HC data set \citep{Huertas-Company+2011} consists of 699684  galaxies from the SDSS-DR7 spectroscopic sample \citep{Abazajian2009}
 with redshift $z\le0.25$ 
 and observed $r$-band magnitude $m_{r}\le18$ mag. 
 \citet{Huertas-Company+2011} provide a morphological classification of the SDSS-DR7 galaxies through a supported-vector-machine  (SVM) classifier \citep[e.g.,][]{Hastie2010}, which was trained with a sub-sample of 2253 SDSS visually-classified  galaxies from \citet{Fukugita2007}.
 The SVM classifier provides a score/probability for a galaxy to be
 either elliptical-lenticular (E/S0) or spiral (S).
  
The second galaxy catalog is the NYU-VAGC  \citep{Blanton2005}. The NYU-VAGC provides the Star Formation Fraction (SFF) of 2506754 SDSS-DR2 galaxies  (see equation \ref{eq:1} for the SFF definition) estimated over the last 300 Myr.
We use the SFF value to divide the galaxies into two  classes:  galaxies with low-to-moderate SFF, and galaxies with high SFF. Quantitative details on the two classes  are provided in  \S\ref{RFNYU}. In the following two sections we train and test two RF classifiers with these data catalogs (PS1-DR2, HC and NYU-VAGC).

%%%%%%%%%%%%%%%%%%%%%%%%%%%%%%%%%%%%%%%%%%%%%%%%%%%%%%%%%%%%%%%%%%%%%%%%

\section{Random forest model for sources in PS1-DR2 based on the HC catalog }
\label{preprocess}
In this section, we analyze the  PS1-DR2  and HC catalogs. We start by crossmatching the two catalogs. For each common source we retain features that are relevant to our subsequent analysis (\S\ref{ffs}).
We then train and test an RF algorithm that classifies the galaxies as E/S0 or S
(\S\ref{TT}, \S\ref{res}).

%-------------------------------------------------------------------------
\subsection{Feature Selection and Pre-processing}
\label{ffs}

As a first step, we identify   sources that are common to  HC  and PS1-DR2 by utilizing a $0.8\arcsec$ search radius. This search radius is optimized to account for a slight difference in astrometry between the two catalogs, without introducing a significant number of spurious associations. \cite{Tachibana2018}, for example, use a similar matching radius for the PS1-DR1 catalog. Furthermore, we discard  PS1-DR2 sources with missing data.
After cross-matching, the data set  consists of 659460 galaxies. 
For each common source, we identify which properties of those listed in   PS1-DR2  are the most relevant to our analysis, and associate the labels (E/S0  or S) from the HC catalog. Specifically, for each PS1 photometry filter, we identify the meaningful features as  the PSF-flux, the Kron-flux \citep{Kron+1980}, and the second moment of the radiation intensity, defined as $<XY>=\int_{Sxy} I(u,v) \,du\,dv$ or $<X^2>=\int_{Sx} I(u) \, du$, where $I$ is the radiation intensity. These properties (which constitute features for our machine learning tools of \S\ref{TT})  are listed in the ``StackObjectAttributes'' table of the PS1-DR2 catalog as \texttt{PSFFlux, KronFlux, momentYY, momentXY and momentXX}.
%-------------------------------------------------------------------------

The features of our training set (i.e. fluxes and moments of the radiation intensity) are distance dependent and, hence, not properly suitable for a machine learning algorithm. We thus engineer the fluxes and moments into a series of features that are not dependent on distance. Specifically, we consider the ratio between the  fluxes and the ratio  of the moments of the radiation intensity for different filters to be meaningful features. 
 This procedure leads to 126 features associated to each of the 659460  sources in our sample. Finally, we standardize the features according to the formula $X_{st}=(X-\mu)/\sigma$, where $X$ is the input feature, while $\mu$ and $\sigma$ are the mean and the standard deviation of the sample, respectively.

The final step of pre-processing is the reduction of dimensionality. To this aim we use the Principal Component Analysis (PCA, e.g., \citealt{Hastie2010}), which performs a linear transformation of a set of correlated features into linearly uncorrelated variables. This algorithm  returns a set of eigenvectors with associated eigenvalues, where larger eigenvalues are associated to eigenvectors that describe the  most of the variance of the data set. The dimensionality reduction is achieved by discarding the components with smaller eigenvalues. We retain the 55 (of the original 126) features responsible for $99.8\%$ of the sample variance.

Before applying machine learning algorithms, we address the potential problem of class imbalance, which occurs when the classes of objects identified by the labels (in our case E/S0 vs.\ S) contain markedly different numbers of elements.  In the HC data set $\approx 39\%$ of the sources belong to the E/S0 class, while $\approx 61\%$ belong to the S class. Since the number of objects is large ($659460$),
and the data set is not heavily imbalanced, we apply standard undersampling, leading to a final balanced data set of 514288 sources.

%-------------------------------------------------------------------------

\subsection{Machine learning with Random Forest}  
\label{TT}
Supervised machine learning classifies objects by learning  a mapping function from the training set and then applying 
the mapping function to previously unseen data. 
A large variety of machine learning algorithms are known, and some have been used in the astronomical literature as well (e.g., \citealt{Dieleman+2015,Lochner+2016,Baldeschi+2017,Baldeschi+2017B,Schanche+2019,Ntampaka+2019,Walmsley2019,Carrasco-Davis2019,Muthukrishna+2019,Villar+2019,Steinhardt2020,Margalef-Bentabol2020,Sravan2020}). 
We apply three algorithms: random forest (RF), supported vector machine (SVM), and boosting  \citep[e.g.,][]{Hastie2010}. The algorithm that leads to the highest classification accuracy is the RF, which we employ below.

The architecture of the RF construction depends on several hyperparameters. Here we adopt the following hyperparameters: (i) the number of estimators, i.e. the number of trees, is fixed at 100; (ii) the maximum depth of a tree can be 10, 12, 14, 16, 18, 20, or 22; (iii) the minimum number of samples required to split an internal node is 2, 3, or 4; (iv) the minimum number of samples required to be a leaf node is 2, 3, or 4. Finally we adopt the Gini\footnote{See \citet[][]{Hastie2010} for a
description of the Gini information gain.} information gain to measure the quality of a split. For each combination of parameters an RF model is produced and then tested using a standard k-fold cross-validation procedure \citep[e.g.,][]{Hastie2010} with $k=4$.  Our results are described in \S\ref{res}.

%------------------------------------------------------------------------
\subsection{Main Results: classification reliability, importance of features and redshift dependence}
\label{res}
\begin{figure*}
\hskip -2 cm
\includegraphics[scale=0.52]{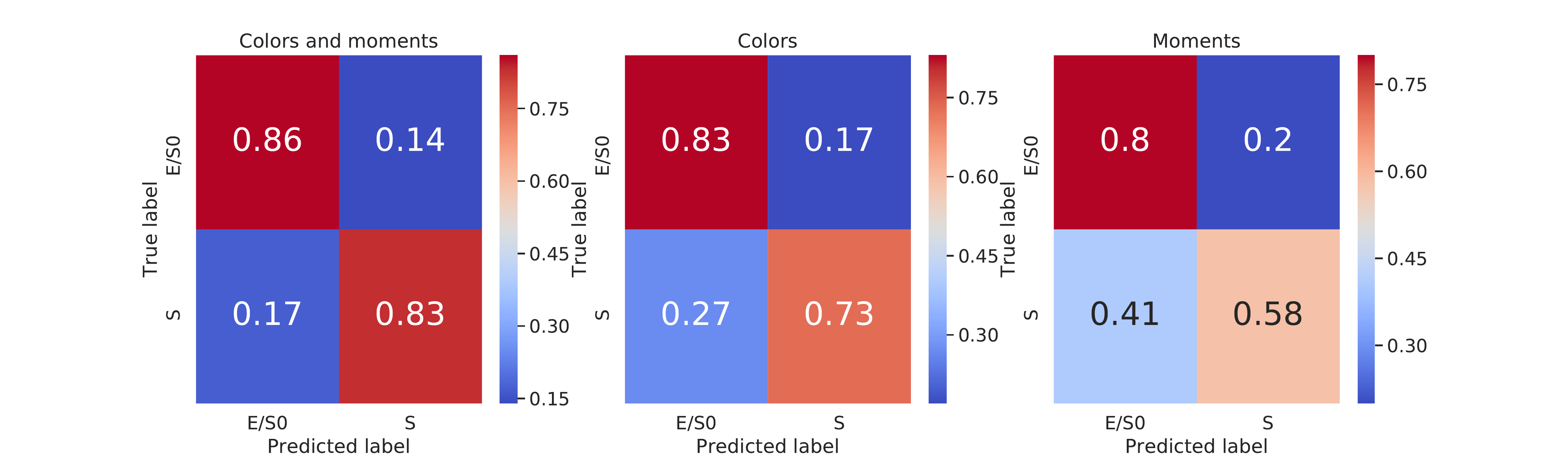}
\caption{Normalized confusion matrix for the two classes of sources: Elliptical-lenticular ($E/S0$) and spiral ($S$). Here we use a random forest (RF) algorithm to classify the objects in the cross validation set of the PS1-DR2-HC dataset. We train the RF classifier by retaining as predictive features colors and the ratio of moments of the radiation intensity (left panel), only colors (middle panel), only the ratio of moments (right panel).
}
\label{fig:conf}
\end{figure*}

The RF in the cross-validation set results in a classification accuracy of $85\%$, meaning that we can reliably infer the morphological properties of galaxies as E/S0 or S by utilizing only the colors (i.e. the flux ratios) and the moments ratio of the radiation intensity.
Figure \ref{fig:conf} (left panel) displays the confusion matrix for the two  classes, from which we conclude that 
the RF model performs equally well for both classes. The data set is balanced and the fraction of sources that are correctly classified is 0.86 and 0.83 for the E/S0 and  S types, respectively.

\begin{figure}
\centering
\includegraphics[width=0.5\textwidth]{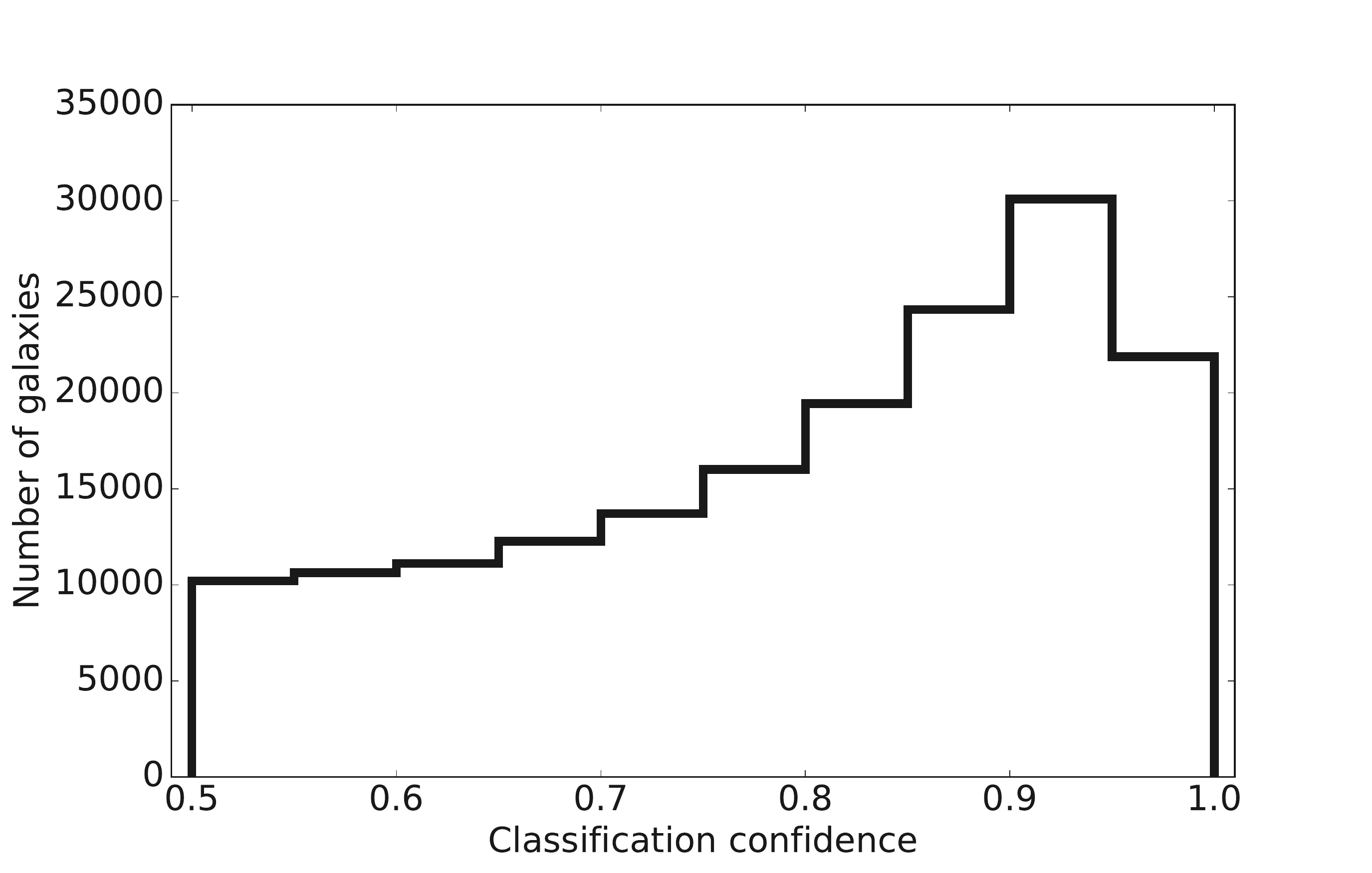}
\caption{Classification confidence distribution for the HC galaxies of being 
 elliptical-lenticular ($E/S0$) or spiral ($S$).
 We define the classification confidence as the classification probability if the classification probability is $>$0.5 and as 1-(classification probability) if the classification probability is $<$0.5.
The distribution has mean and median values of $0.80$ and $0.83$, respectively.
}
\label{fig:3}
\end{figure}

The RF classifier gives the probability that a source belongs to either the E/S0 or the S class. The predicted classification probabilities of an input sample are computed as the mean predicted class probabilities of the trees in the forest. The class probability of a single tree is the fraction of samples of the same class in a leaf. Figure \ref{fig:3} shows the distribution of the classification confidence for the sources in the test set to belong to one of the two classes (E/S0 or S). The distribution is bound between 0.5 and 1, and has a mean value of 0.8 and a median value of 0.83. We define the classification confidence as equal to the classification probability if this is $>$0.5, otherwise we define it as 1-(classification probability). We note that $74\%$ of the sources  in the HC sample have a classification confidence (of belonging to the E/S0 or the S class) that is larger than $70\%$. This suggests that the classifier provides reliable classifications in most cases.

 The importance of the features in an RF model is typically measured  by estimating  the increase in the model's predictive accuracy after permuting the feature \citep[see e.g.,][]{Hastie2010,Fisher2018}. A feature is considered important if shuffling its values decreases the model accuracy, which implies that  the model relied on that specific feature for the prediction. We  estimate the importance of the colors and the ratio of  moments of the radiation intensity in the morphological classification of galaxies as follows.  We build two models, where one model  considers only the ratio of moments as to be features, and the other considers only the colors to be features. We find a classification accuracy of $78\%$ for the RF model that retained only the colors as features, while a classification accuracy of $69\%$ is achieved by retaining only the ratio of the moments. Figure \ref{fig:conf}  displays  the confusion matrix for both the colors (middle panel) and moment classifiers (right panel), respectively. We conclude that while
 colors are the most significant features, the combination of colors and ratio of moments leads to 
an improved final classification accuracy.

\begin{figure}
\centering
\includegraphics[width=0.5\textwidth]{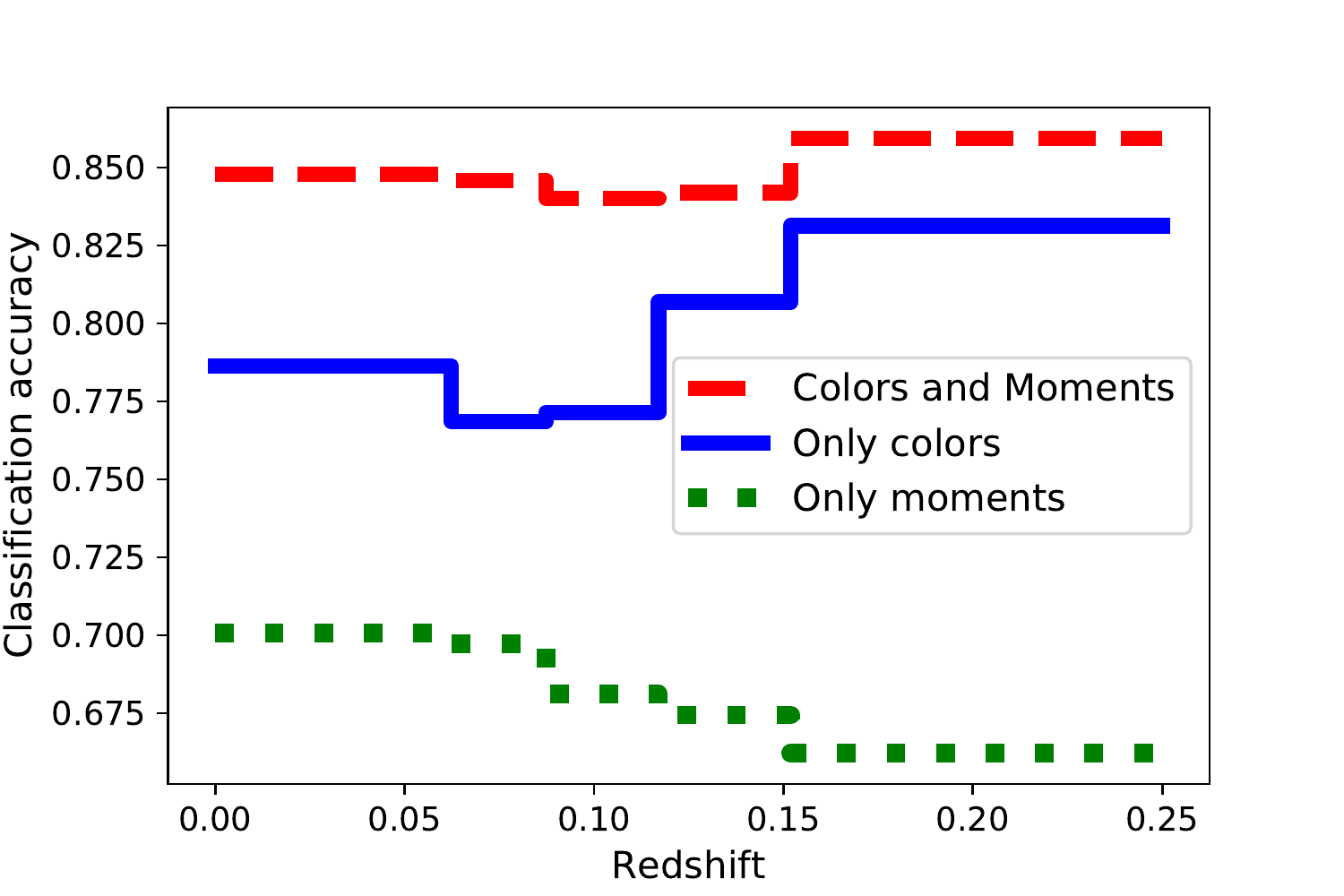}
\caption{Classification accuracy as a function of the galaxy redshift for three RF models for  the PS1-DR2-HC dataset. Red dashed line: both colors and ratios of moments of the radiation intensity are used as input features for the RF classifier. Blue line: only colors are used as input features for the RF classifier. Green dotted line: only the ratios of moments of the radiation intensity are used as input features for the RF classifier.
The figure reveals that the classification accuracy is approximately constant for $z<0.25$ when the RF is trained using colors and ratios of the radiation intensity as features.
Data at $z>0.15$ have been collected into a single bin due to limited statistics.
}
\label{fig:5}
\end{figure}

Finally, in Figure \ref{fig:5} we explore the dependence of the classification accuracy on the redshift of the sources.

When colors and moments of radiation intensity are retained as features, (red dashed line in Figure \ref{fig:5}), the classification accuracy is approximately constant with redshift, suggesting that our methodology provides results that are not heavily distance dependent at $z<0.25$ and are mostly unbiased with distance. 
When colors are the only features, (blue line in Figure. \ref{fig:5}), the classification accuracy is slightly larger at higher redshift, while when we use the ratio of moments of the radiation intensity as only features of the model (green dotted line in Figure \ref{fig:5}), 
the classification accuracy decreases at higher redshift. From Figure \ref{fig:5} it is also clear that the classification accuracy of the RF classifier that retains both colors and ratios of moments of the radiation intensity is significantly higher at all redshifts.

%%%%%%%%%%%%%%%%%%%%%%%%%%%%%%%%%%%%%%%%%%%%%%%%%%%%%%%%%%%%%
\section{Random forest model for sources in PS1-DR2 based on the NYU-VAGC}
\label{RFNYU}

In this section, we train and test an RF classifier  by combining the features in  PS1-DR2 with the galaxies star formation properties derived from the NYU-VAGC. The goal is to build an RF model that can classify sources in the PS1-DR2 into two distinct classes consisting of galaxies with a low-to-moderate  star formation fraction (SFF, \S\ref{lab_}), and galaxies with  high SFF.

\subsection{Selection of Sources and definition of Labels  for the NYU-VAGC}
\label{lab_}
The NYU-VAGC consists of $2506754$ sources.  Among these, we select sources with redshifts in the range $0.002\le z \le 0.5$ to limit 
the contamination by stars in the Galaxy and lower-quality observations.
We discard sources with bad photometry (e.g., missing data and high uncertainty in the fluxes).
These cuts result in a catalog of $662804$ sources  that are used in the following analysis.

The NYU-VAGC catalog provides the  star formation fraction (SFF\footnote{The SFF parameter was named as B300 in \cite{Blanton2007}.}) for each source 
 defined as \citep{Blanton2007}:
\begin{equation}\label{eq:1}
SFF \equiv \frac{\int_{t_0-0.3\,Gyr}^{t_0}  SFR(t) \, dt}{\int_{0}^{t_0} SFR(t) \, dt},
\end{equation}
where $t_{0}$ is the present epoch and  $SFR(t)$ is the star formation rate as a function of time. $SFF$ is thus a continuous unitless variable that represents the  fraction of  star formation that has occurred in the past 0.3 Gyr.
We use the RF classifier developed in \S\ref{preprocess} that retains both colors and ratios of moments of the radiation intensity as input features, to divide the NYU-VAGC sources into two SFF classes based on their early-type (i.e. $E/S0$) or late-type (i.e. $S$) classification (i.e. to convert the continuous SFF variable above into a discrete feature that can be used as a label).

\begin{figure}
\centering
\includegraphics[width=0.5\textwidth]{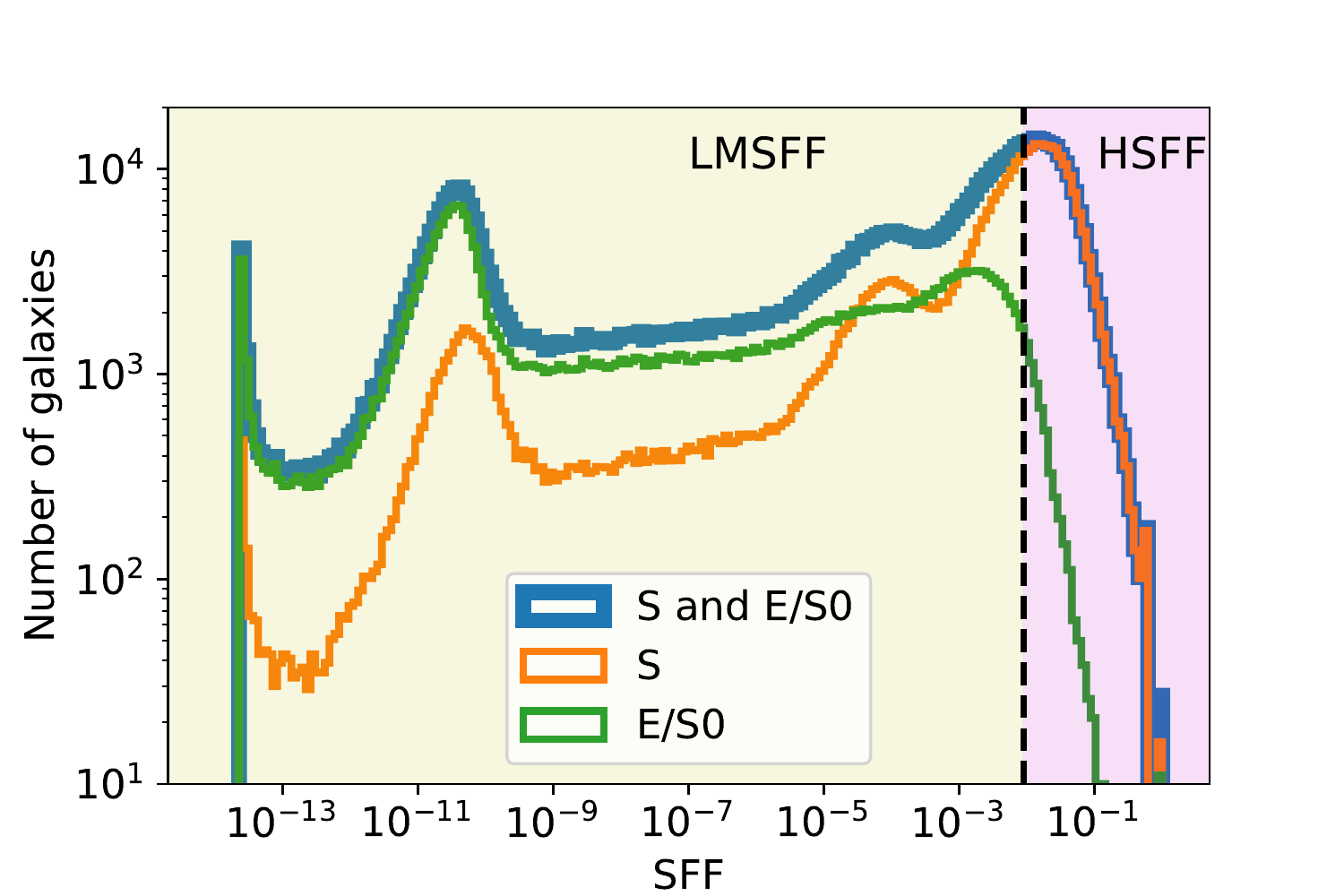}
\caption{SFF distribution of NYU-VAGC sources. Green: elliptical-lenticular (E/S0) sources. Orange: spiral (S) sources.
Blue: complete sample of E/S0 and S sources. We perform object classification (E/S0 vs.\ S)  using the RF classifier developed in  \S\ref{TT}. We labeled the sources on the left of the vertical black dashed line as LMSFF and sources on the right of the vertical black line as HSFF.
Most of the HSFF sources are S, while the LMSFF sources are both S and E/S0 (mainly E/S0).
}
\label{fig:6}
\end{figure}

Figure \ref{fig:6} displays the distribution of SFF of galaxies in the NYU-VAGC classified as $E/S0$ (green line) and $S$ (orange line). Figure \ref{fig:6} reveals that most of the $E/S0$ galaxies  are associated with a low-to-moderate SFF, while most of the $S$-classified galaxies are associated with larger SFFs, as expected.  In particular, for $SFF\gtrsim 10^{-2}$  most sources are late-type galaxies, while for $SFF\lesssim10^{-2}$ the galaxy types consist of a mixture of both early-type and late-type (where the early-type galaxies are predominant). Therefore,  we  split the population of galaxies into two distinct classes: for $SFF>9\times10^{-3}$, the sources are labeled as galaxies with high  star formation fraction (HSFF), whereas for smaller values ($SFF<9\times10^{-3}$) the sources are labeled as galaxies with low-to-moderate  star formation fraction (LMSFF).

\begin{figure}
\centering
\includegraphics[width=0.5\textwidth]{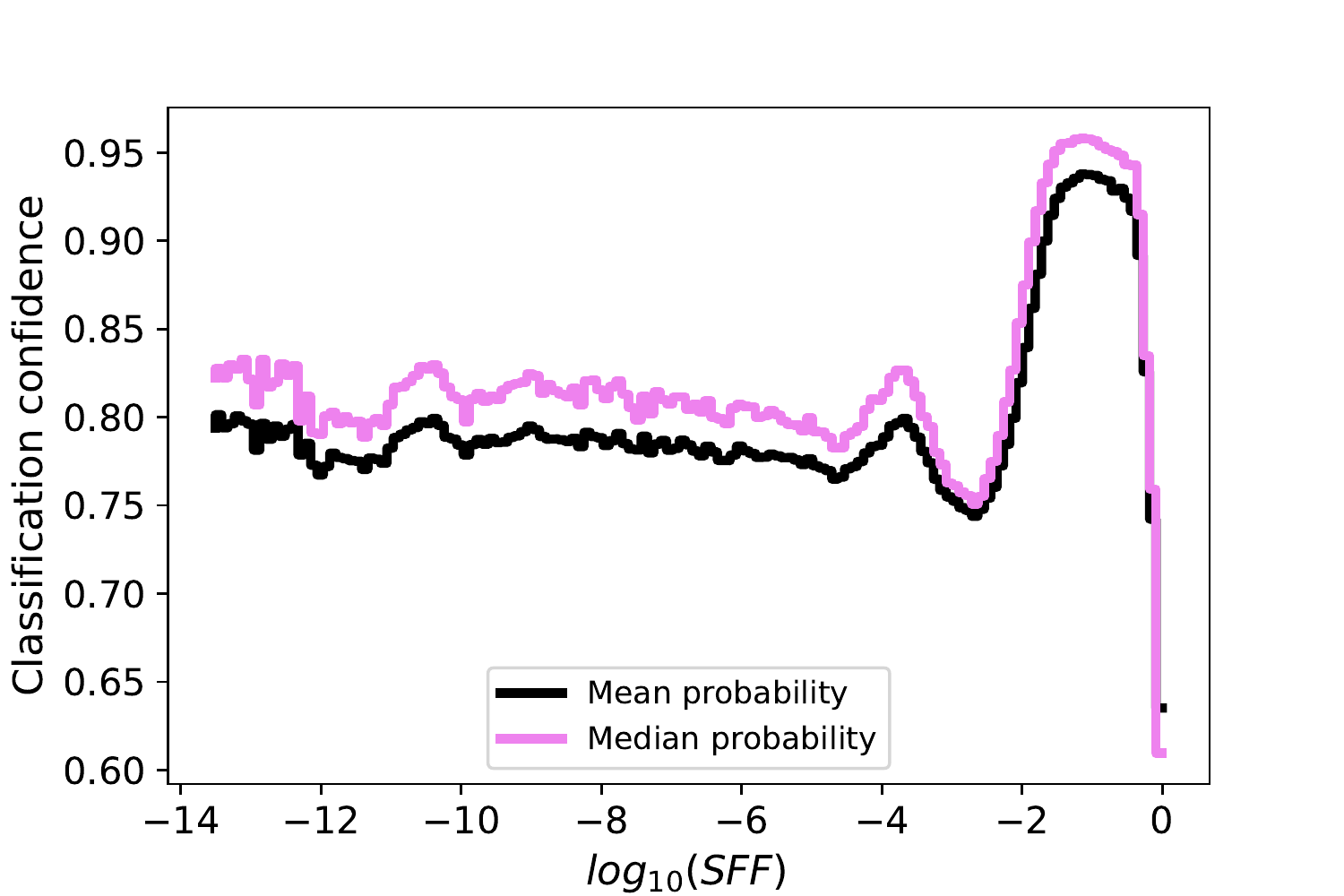}
\caption{
Mean (black) and median (pink) classification confidence for each of the SFF  distribution bins in the NYU-VAGC catalog. Galaxies with very large SFF$>10^{-2}$ are classified with higher confidence.
}
\label{fig:10}
\end{figure}

We can also estimate the mean and median classification confidence of the NYU-VAGC galaxies to be either $E/S0$ or $S$ for each bin of the  SFF distribution, and we show the results in Figure \ref{fig:10}.  The classification confidence is estimated by employing the RF classifier developed in \S\ref{TT}.
Figure \ref{fig:10} reveals that the median classification confidence lies between 0.8 and 0.85 for $SFF<10^{-4}$, decreases to 0.75 at $SFF\approx 10^{-3}$ and increases above 0.9 at  $SFF\approx10^{-1}$.
HSFF galaxies are thus classified with higher confidence.

%--------------------------------------------------------------------------
\subsection{Pre-processing, Training, Testing and Results}
\label{SubSec:42}
In \S\ref{lab_} we defined the labels  for the classification process (HSFF vs LMSFF).
Here, we follow the methodology developed
in \S\ref{preprocess}  (i.e. pre-processing, training, and testing by cross-matching sources in the PS1-DR2 and HC data-set) for the  NYU-VAGC catalog. 
First, we select common sources  between  NYU-VAGC and PS1-DR2  by crossmatching the two catalogs using a $0.8\arcsec$  radius.
We  only consider colors as predictive features for the training/testing set, as considering colors and ratios of moments of the radiation intensity would lead to a lower  classification accuracy (see the discussion below).
 The features are then standardized, and a PCA is performed to reduce the dimensionality of the data set as described in \S\ref{preprocess}.
 We start with 25 features (the ratio of fluxes in different photometric bands). The PCA reduces the number of meaningful features to 16
 responsible for $99.7\%$ of the sample variance.

This procedure results in a data set where $\approx 25\%$ of sources belong to the HSFF class, while $\approx 75\%$ belong to the LMSFF class. We adopt undersampling to balance the data set,  resulting in a final data set of $323220$ sources. After data standardization and balancing, we train and test an RF using 4-fold cross-validation. The RF classifier is optimized employing  the grid-search method discussed in \S\ref{TT}. The RF classifier  reaches a classification accuracy of $89\%$ in the 4-fold cross-validation.

\begin{figure}
\centering
\includegraphics[width=0.5\textwidth]{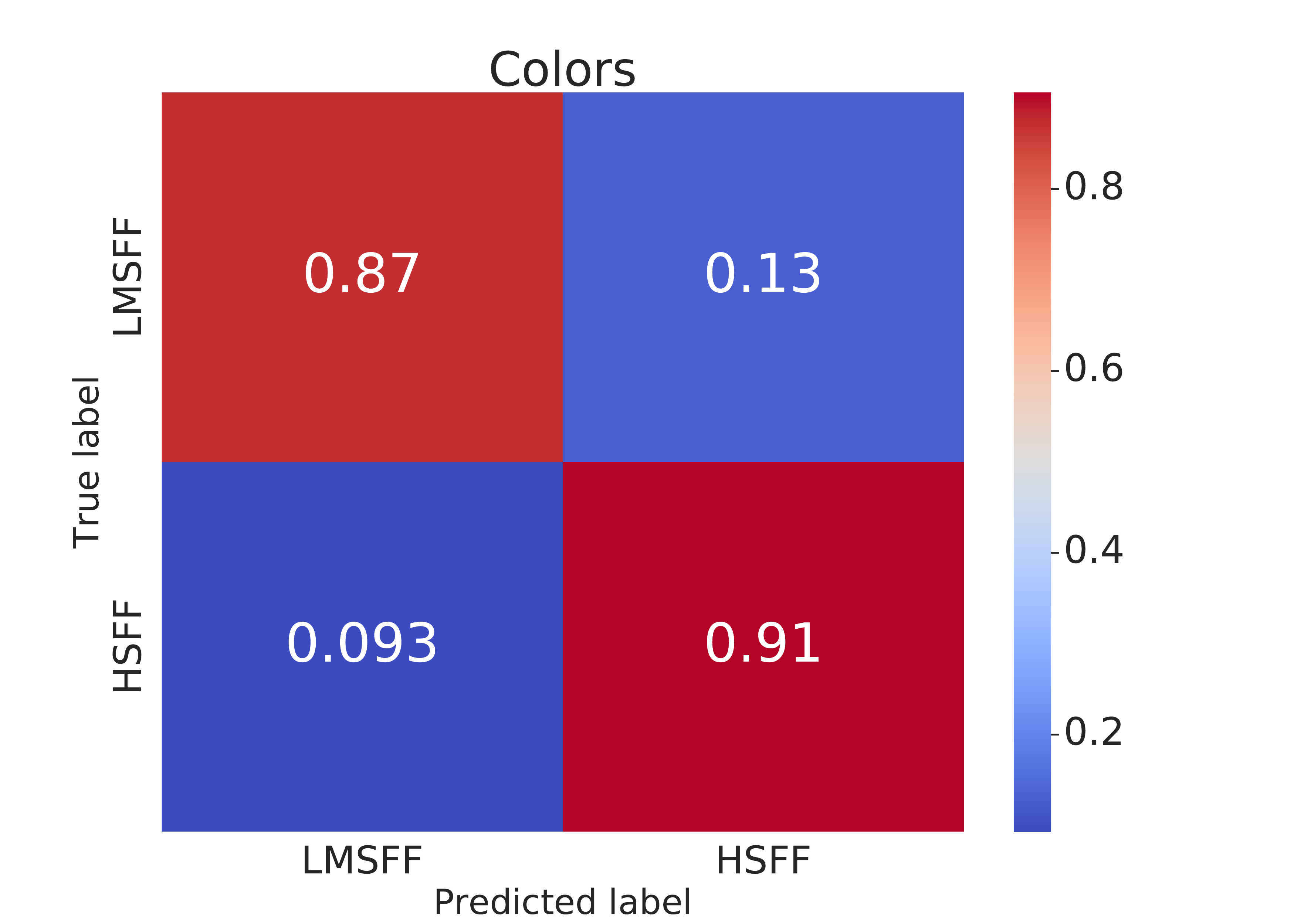}
\caption{Normalized confusion matrix for the two classes HSFF and  LMSFF of the NYU-VAGC catalog.  
Only colors were used as input features for the RF classifier described in \S\ref{SubSec:42}.
}
\label{fig:8}
\end{figure}

\begin{figure}
\centering
\includegraphics[width=0.5\textwidth]{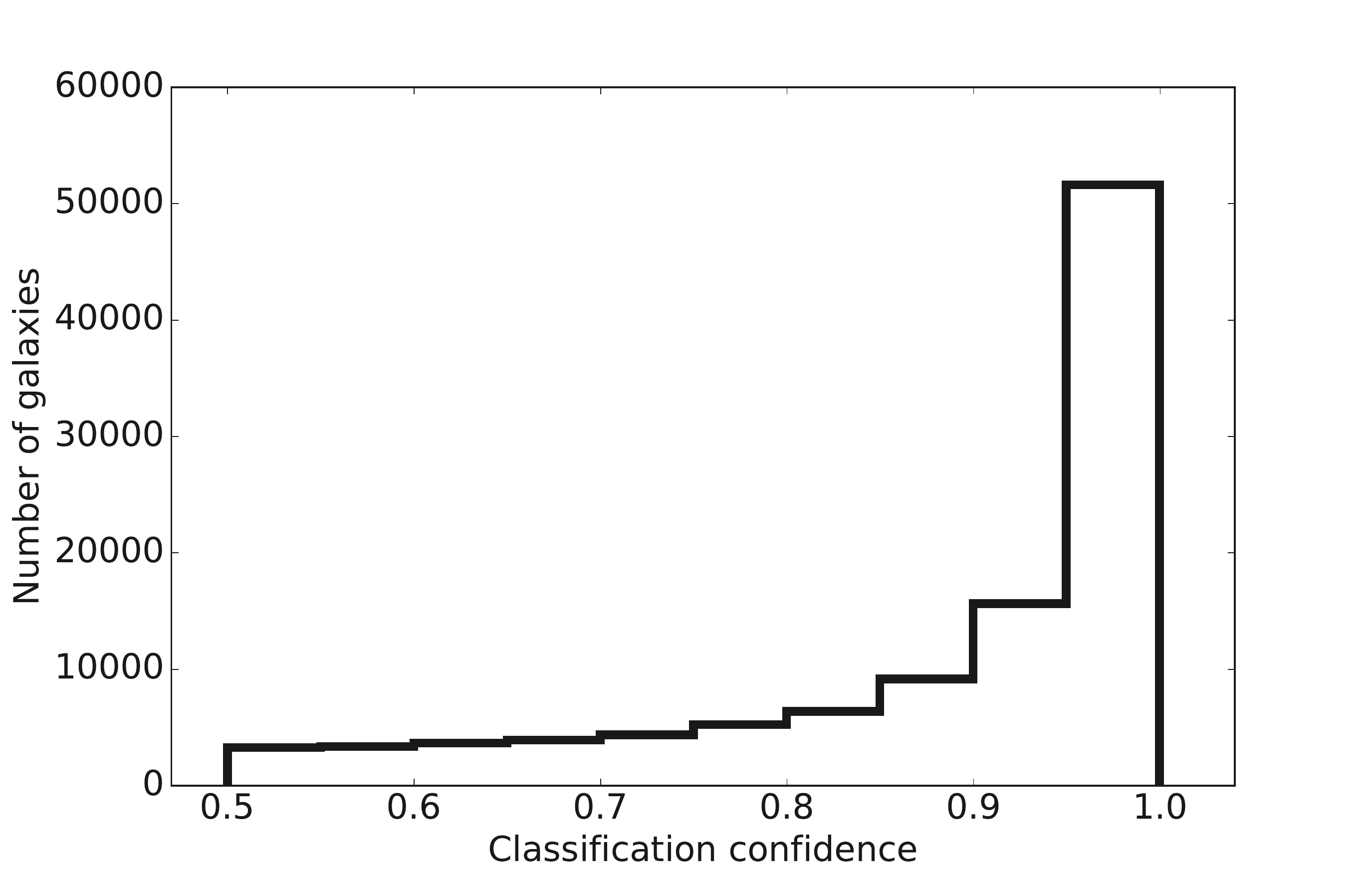}
\caption{Distribution of the classification confidence of the NYU-VAGC galaxies labeled as HSFF or LMSFF. The probability that an individual galaxy is HSFF or LMSFF is calculated by employing the RF classifier developed in \S\ref{RFNYU}. The distribution has mean and median values of 0.88 and 0.94, respectively. }
\label{fig:7}
\end{figure}

\begin{figure}
\centering
\includegraphics[width=0.5\textwidth]{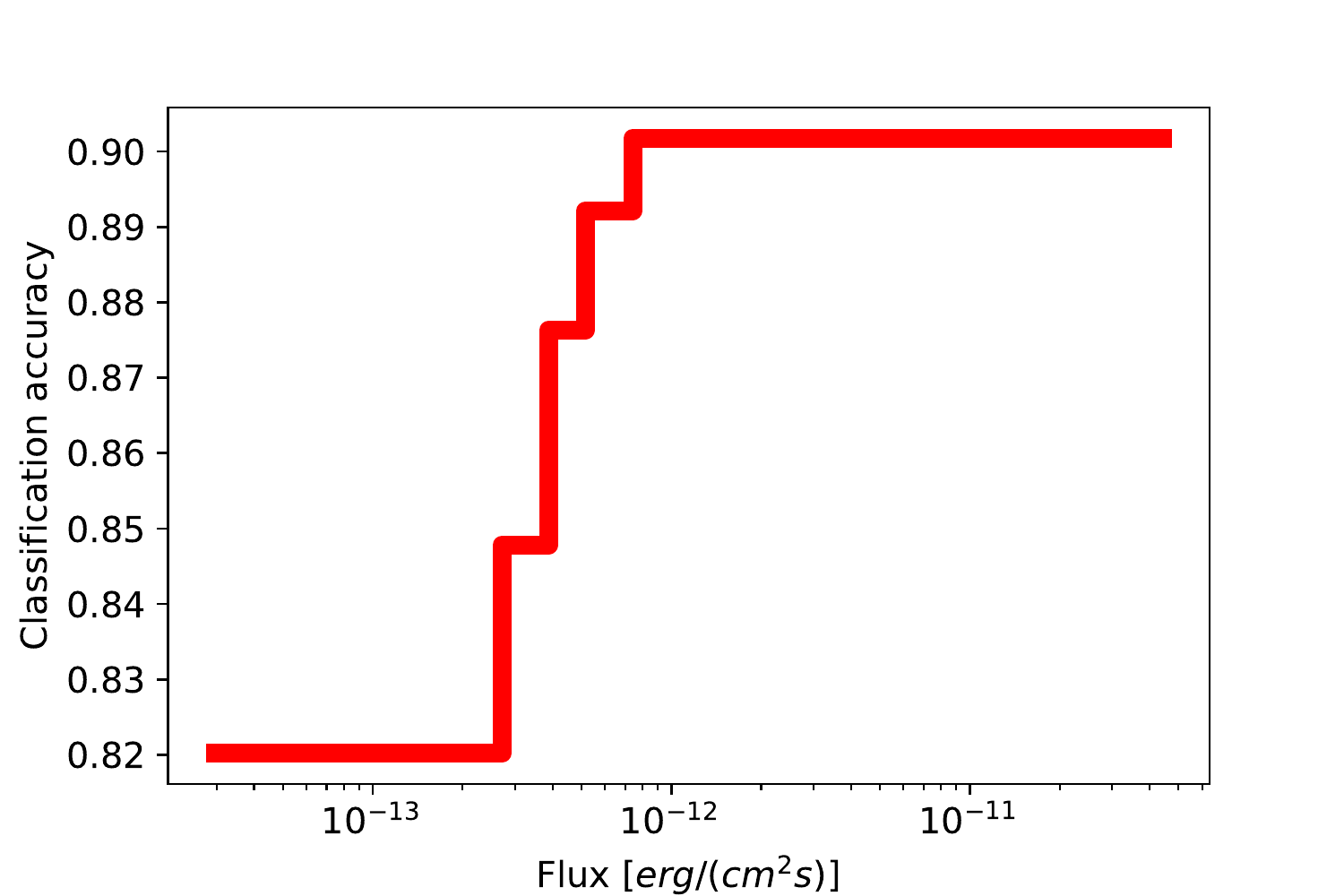}
\caption{Classification accuracy (HSFF and LMSFF) vs.\ integrated flux 
of the NYU-VAGC galaxies. Each bin in this histogram contains the same number of sources. As expected, brighter sources are easier to classify. 
}
\label{fig:99}
\end{figure}

Figure \ref{fig:8} shows the confusion matrix for the HSFF vs.\ LMSFF classification. We find that the fraction of sources that are correctly classified is 0.91 and 0.87 for the HSFF and  LMSFF types, respectively. In Figure\ref{fig:7} we display the distribution of the classification confidence  for each source in the test set. 
The distribution is bound between 0.5 and 1, and the mean and median values of the distribution are 0.88 and 0.94, respectively. These results suggest that  the RF model can predict the label of the galaxy (HSFF or LMSFF) with a reasonably high probability by using the colors as the only meaningful features.

Although the high classification accuracy within the cross validation set is a reliable measure for estimating the validity of the RF model, it is also important to explore the dependence of the classification accuracy on the brightness of the galaxies. In Figure \ref{fig:99} we display the classification accuracy as a function of the integrated flux in the five PS1 filters. The integrated flux was estimated by computing the trapezoidal approximation of the spectral energy distribution. The figure shows that our classification accuracy  is a monotonically increasing function of the source flux, as expected (i.e. brighter sources are easier to classify).

\begin{figure}
\centering
\includegraphics[width=0.5\textwidth]{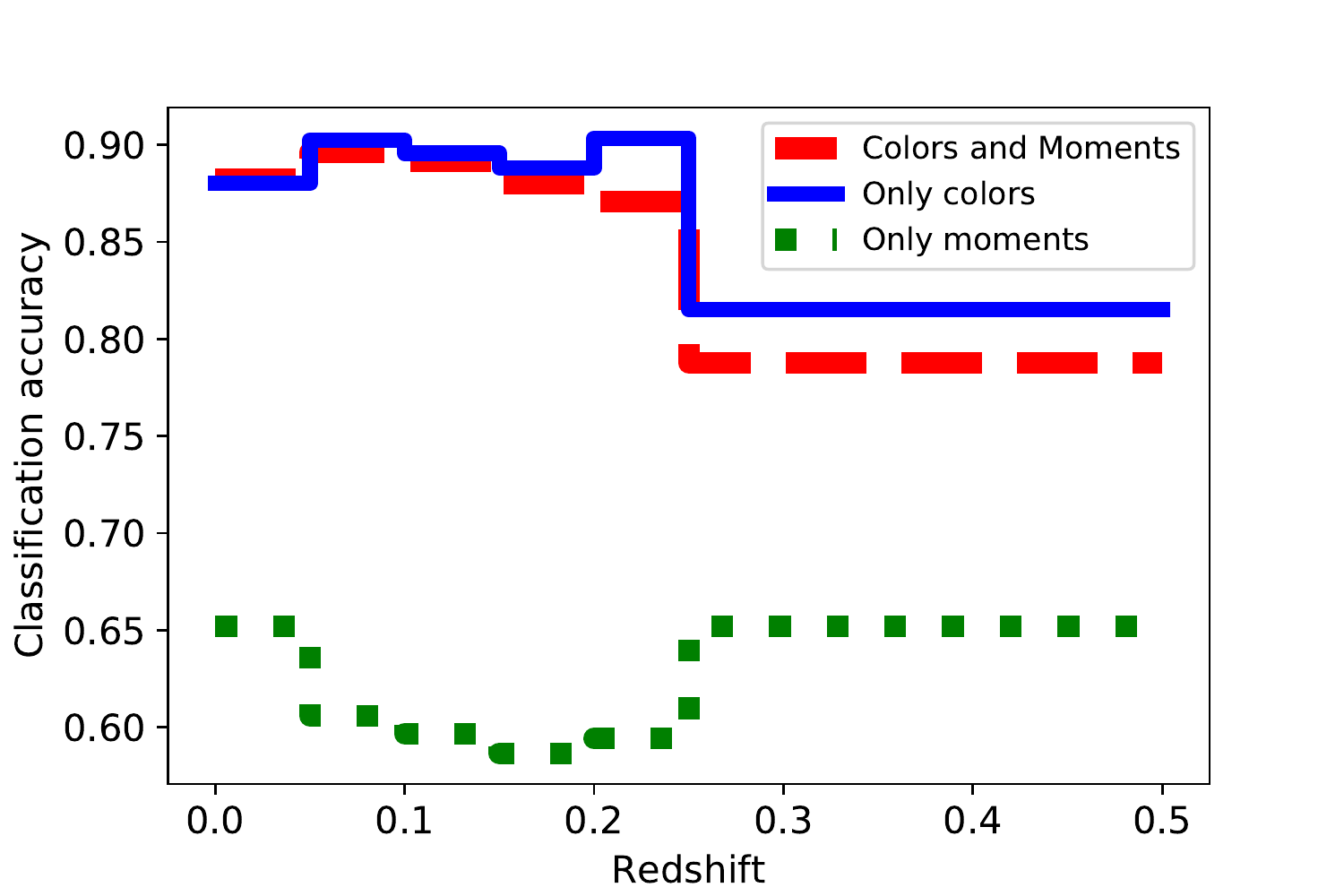}
\caption{Classification accuracy (HSFF vs.\ LMSFF) as a function of redshift for NYU-VAGC galaxies.
Dashed red line: both colors and ratios of moments of the radiation intensity have been used as input features for the RF classifier. Blue  line: only colors as input features.
Green dotted line: only the ratios of moments as input features. Data at $z\ge0.25$ have been collected into a single bin due to limited statistics.
} 
\label{fig:9}
\end{figure}

In \S\ref{res} and Figure \ref{fig:5} we demonstrated that the E/S0 vs.\ S galaxy classification is most accurate when using both colors and ratios of moments of the radiation intensity as features, and we explored the dependency of the classification accuracy on the sources redshift. Here, we reproduce the exercise of  \S\ref{preprocess} for this second RF classifier. Figure \ref{fig:9} illustrates the dependence of the classification accuracy on redshift for each of the three choices of training features. The figure clearly demonstrates that the highest classification accuracy is reached by considering  only colors as input features. The classification accuracy is approximately constant up to $z \approx 0.25$
and decreases at larger values.

\begin{figure}
\centering
\includegraphics[width=0.5\textwidth]{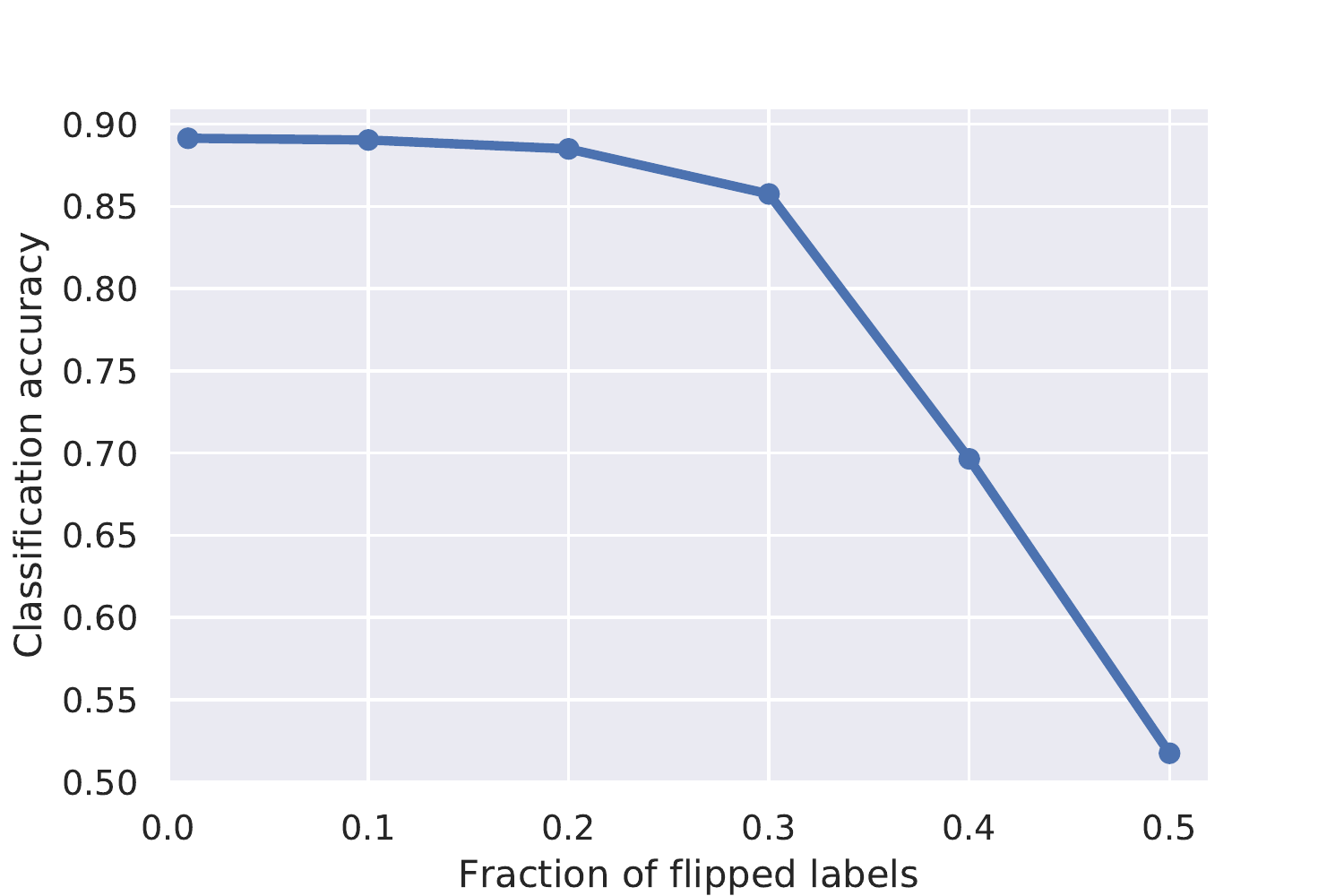}
\caption{Classification accuracy in the test set (HSFF and LMSFF) vs.\ 
fraction of flipped labels in the training set in  the NYU-VAGC dataset.
The figure shows that the RF model is robust, as the classification accuracy 
is $>85$\% even when $30\%$ of the labels are flipped.
}
\label{fig:flip}
\end{figure}

All these results implicitly assume  that the training/testing labels are, indeed, accurate. In fact, the labels are likely to be inaccurate: a galaxy that is labeled as having HSFF may actually have LMSFF. This is due to the fact that the SFF is a difficult  parameter to estimate and is subject to some degree of uncertainty.
Therefore, an unknown fraction of galaxies may be mislabeled. Even though a  fraction 
of galaxies may be mislabeled, our RF algorithm  is robust to the presence of partially mislabeled data. 
A way to assess the robustness of the model is to randomly flip a fraction of labels  in the training set and then estimate the classification accuracy. In Figure \ref{fig:flip} we show the classification accuracy in the test set versus the fraction of flipped labels in the training set. The figure reveals that even if $30\%$ of the labels are flipped the classification accuracy is still very high ($\approx 85\%$). For  larger fractions of flipped labels, the classification  accuracy approaches $50\%$ (equivalent to random guessing), as expected.

We end with a consideration on the performance of ``lighter'' models that are based on a significantly smaller number of features, and yet reach interesting levels of accuracy. 
In this section we trained a large model with 16 meaningful features, and reached a classification accuracy of 0.89 in the cross validation set. It is also possible to train and test a much simpler model that uses only 2 features  and still obtain reliable results. If we train an RF model with two features (i.e. the Kron-flux ratio between the $g-$ and $r-$band filter $F_{K,g}/F_{K,r}$, and the $i$ and $z$ bands, $F_{K,i}/F_{K,z}$), we obtain a classification accuracy of 0.86 in the cross validation set. Using  PSF fluxes, the RF model achieves a classification accuracy of 0.81. Therefore, even with a  highly simplified model, we obtain  classification accuracies that are comparable to the main RF model of this section.
In the next section we apply the more complex and sophisticated 16-feature RF model developed in this section to the entire PS1-DR2 catalog because of its larger classification accuracy (0.89) and its intrinsic flexibility that allows the model to be applied when either (some of) the Kron or the PSF fluxes are missing for a given PS1-DR2 source.

%%%%%%%%%%%%%%%%%%%%%%%%%%%%%%%%%%%%%%%%%%%%%%%%%%%
\section{Classification of P\lowercase{an}-STARRS Sources based on their star formation properties and morphology}
\label{panst}

In \S\ref{preprocess} we have developed RF models to classify galaxies as E/S0 vs.\ S, while in \S\ref{RFNYU} we have trained an RF classifier that discriminates between HSFF and LMSFF galaxies. In this section, we apply both  classification frameworks  to the entire PS1-DR2 data set  and we  build a catalog with the classification of P\lowercase{an}-STARRS  sources based on their morphology and SFF. Specifically,  we pre-process the  PS1-DR2  features (in this case colors\footnote{In \S\ref{preprocess} we trained three RF models with a different set of features: i) colors and ratio of moments of the radiation intensity; ii) colors only; and iii) ratio of moments of the radiation intensity only. The largest classification accuracy was achieved with the first RF model (0.85), while we achieve a classification accuracy of 0.78 for the second model, and 0.69 for the third model. Despite the larger classification accuracy of the first model, here we employ the second RF model based on colors only as most sources in PS1-DR2 lack measurements of the moments of the radiation intensity.})   through the standard procedure  (scaling and PCA) outlined in \S\ref{preprocess}  and \S\ref{RFNYU} using the PCA models estimated on the HC (\S\ref{preprocess}) and  NYU-VAGC (\S\ref{RFNYU}) catalogs, respectively. We then apply the RF models developed in  \S\ref{preprocess} and \S\ref{RFNYU} to the PS1-DR2 and we present the results of our classification in a catalog that is described in detail in Appendix \ref{catalog_desc}.

\begin{figure}
\centering
\includegraphics[width=0.55\textwidth]{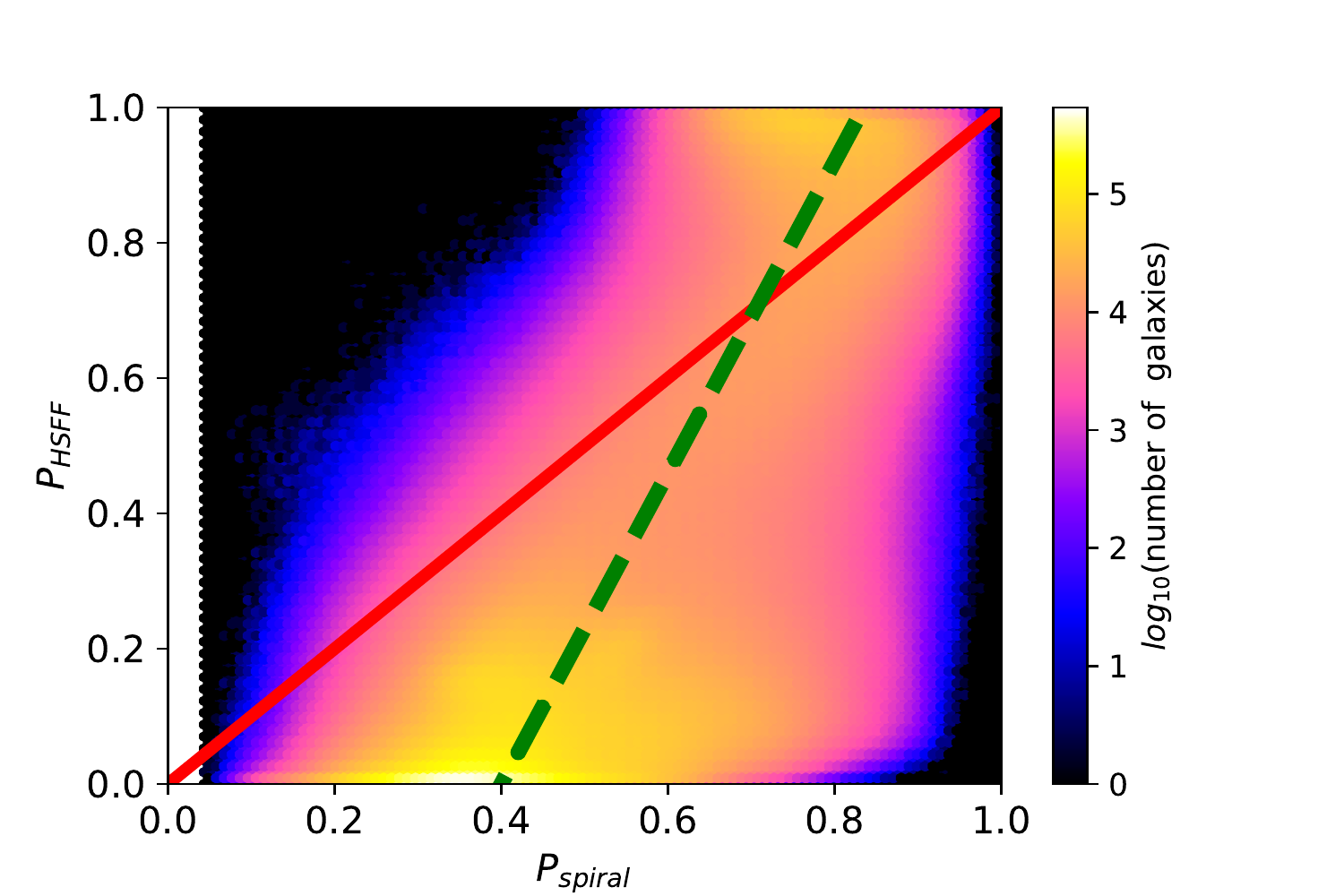}
\caption{ $P_\mathrm{HSFF}$  vs $P_\mathrm{spiral}$  for  PS1-DR2 galaxies (sources with $P_{*}<0.2$).
The red line  identifies  the locus of the plane for which $P_\mathrm{HSFF}=P_\mathrm{spiral}$. The two probability scores show some degree of correlation, as expected (i.e. spiral galaxies tend to have HSFF, while elliptical and lenticular galaxies cluster at  LMSFF). The green dashed line ($P_\mathrm{HSFF}=2.3 P_\mathrm{spiral}-0.9$) represents the approximate correlation between $P_\mathrm{HSFF}$  and $P_\mathrm{spiral}$. The $P_\mathrm{spiral}$ score  never approaches 0 because of an intrinsic difference between the cross-validation
set of the HC catalog and the  PS1-DR2 data set. }
\label{fig:cor}
\end{figure}

The average classification confidence that we obtain for the E/S0 vs.\ S classification is lower than the classification confidence from the HSFF vs.\ LMSFF classifier. The two classifications do however show some degree of correlation, as expected (Figure \ref{fig:cor}), as $S-$galaxies tend to be associated with HSFF sources and $E/S0$ galaxies to LMSFF sources. Figure \ref{fig:cor} shows that the relation between the two indicators is not $P_\mathrm{HSFF}=P_\mathrm{spiral}$ (with scatter), but closer to $P_\mathrm{HSFF}=2.3 P_\mathrm{spiral}-0.9$  (with large scatter), suggesting that sources with  $P_\mathrm{spiral}\le 0.5$ can effectively be mapped into sources with low $P_\mathrm{HSFF}\le 0.25$ (and hence limited star formation).
While we do not necessarily recommend the use of this RF classifier in quantitative studies of galaxy morphology, in \S\ref{transient} we find that $P_\mathrm{spiral}$ can be empirically used in transient surveys to control the purity of core-collapse vs.\ type-Ia SN samples.  

In the rest of this section we  discuss the effects of star-galaxy misclassification, 
brightness,  missing data and Galactic extinction on our results, and we provide the reader with guidelines on how to interpret and use the results from our RF  models (\S\ref{FHSF}-\S\ref{extinction}).

%------------------------------------------------------
\subsection{Effects of Star-Galaxy Misclassifications and Missing Data}
\label{FHSF}

\begin{figure*}
\centering
\includegraphics[width=0.65\textwidth]{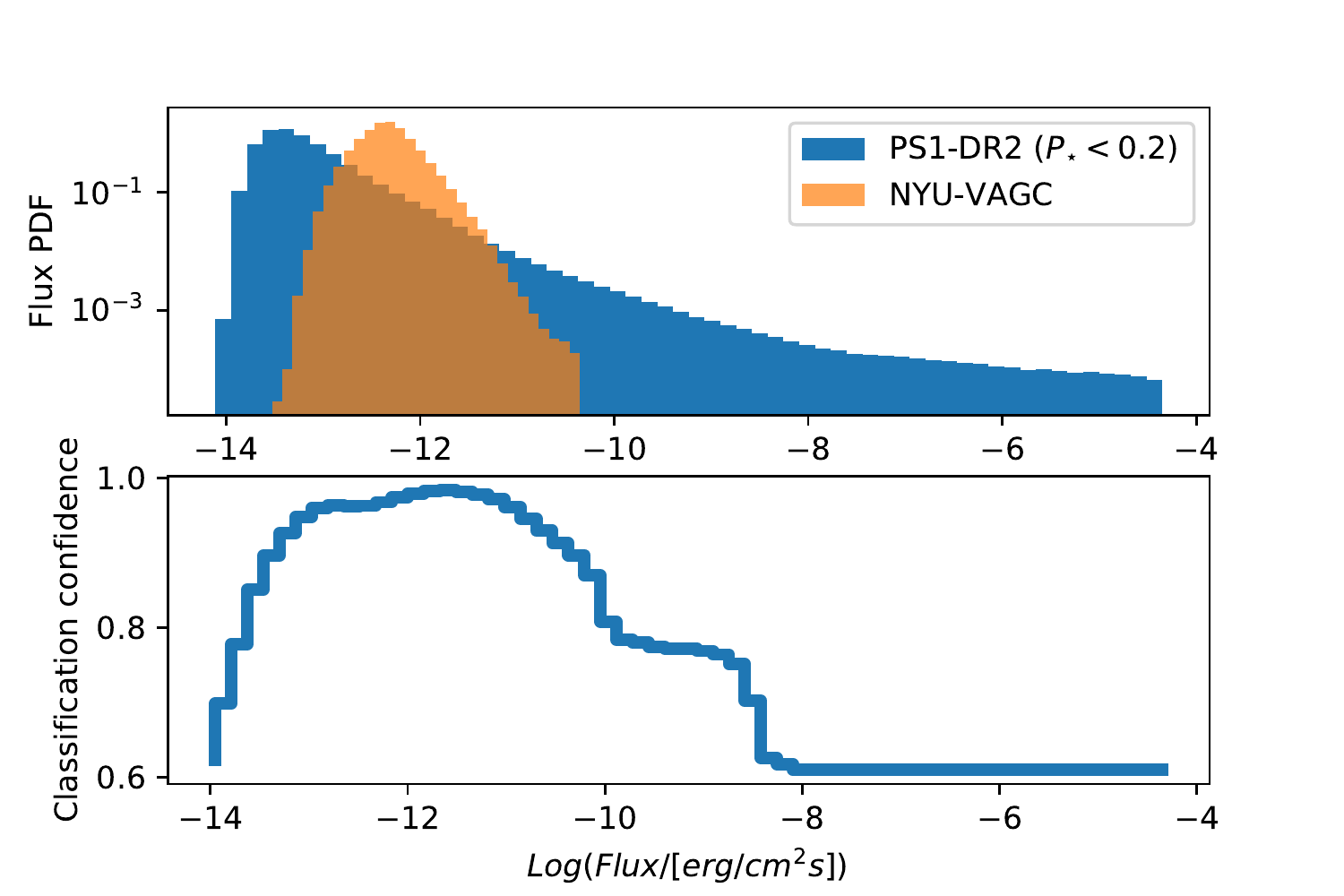}
\caption{\emph{Upper panel}: Empirical probability density function (PDF) of the integrated flux  over the PS1 band-pass 
of galaxies (i.e. sources with $P_{\star}<0.2$) in the PS1-DR2 catalog (blue) and the NYU-VAGC (orange). Log scale is used for the y-axis. 
 \emph{Lower panel}: median classification confidence that  PS1-DR2 galaxies are correctly labeled as HSFF or LMSFF, as a function
of the integrated flux.
 The classification confidence decreases when the flux of the PS1-DR2 galaxies  significantly diverges from the flux of sources in the cross-validation set. 
}
\label{fig:13000}
\end{figure*}

Dissimilarities between the training-testing set and PS1-DR2 include the following: (i) PS1-DR2 contains galaxies 
as well as non-extended objects, i.e. stars; (ii) the training/testing data sets are biased towards brighter sources, implying that the training-testing data sets are only partially representative of sources in the PS1-DR2; (iii) rows with missing data are present in the PS1-DR2 data set, but in the training/testing data set we only considered  sources without missing values.

We address the star-galaxy misclassification issue using the results from \cite{Tachibana2018}. For each P\lowercase{an}-STARRS source these authors provide a score $P_{\star}$ that quantifies the probability that the source is point-like (i.e. a star). We run the RF classifiers
on the entire PS1-DR2 data set irrespective of the $P_{\star}$ value, and for each source we list $P_{\star}$,  the probability of having a high SFF $P_\mathrm{HSFF}$ as derived by our RF classifier in \S\ref{RFNYU}, and the probability of having a spiral galaxy morphology $P_\mathrm{spiral}$   (from \S\ref{preprocess}). Each source in our catalog thus has three separate probability scores, allowing the user to apply a custom cut on $P_{\star}$ as needed. For reference,  $P_{\star}>0.8$ indicates that the object is a star with reasonably high confidence (hence $P_{\star}<0.2$ can be considered highly suggestive of a galaxy-type celestial object). In the remainder of the paper we refer to PS1-DR2 sources with $P_{\star}<0.2$ as ``galaxies''.

The presence of biases between the cross validation data sets and the application set is very common in many machine-learning implementations, and, depending on the degree of bias, cannot be easily mitigated. One  way to visualize the  amount of bias between the cross validation set and the whole  PS1-DR2 catalog  is to compare the distributions of the flux integrated over the PS1 band pass (i.e. $g$- to $y$-band) of sources in the two data sets, which is shown in Figure \ref{fig:13000} (upper panel).  We consider only PS1-DR2 sources with $P_{\star}<0.2$  (i.e galaxies). The median integrated flux of sources in the cross validation set (i.e. the NYU-VAGC) is $\sim$2.6 times larger than the median integrated flux of the  entire PS1-DR2 data set, which also shows a significantly broader distribution.  Indeed, $23\%$ of the PS1-DR2 galaxies have a lower integrated flux than the minimum  value of the  cross validation set, and only  $0.3\%$ of the PS1-DR2 galaxies have a larger integrated flux than the maximum value of the  cross validation set.  In the lower panel of Figure \ref{fig:13000} we display the median classification confidence for the PS1-DR2 galaxies of having HSFF or LMSFF as a function of the integrated flux. This figure reveals that the classification confidence decreases when the integrated flux of the  PS1-DR2 galaxies differs from the values  in the cross validation set, as expected.

\begin{figure}
%\centering
%\vskip -1.8 cm
\includegraphics[width=0.5\textwidth]{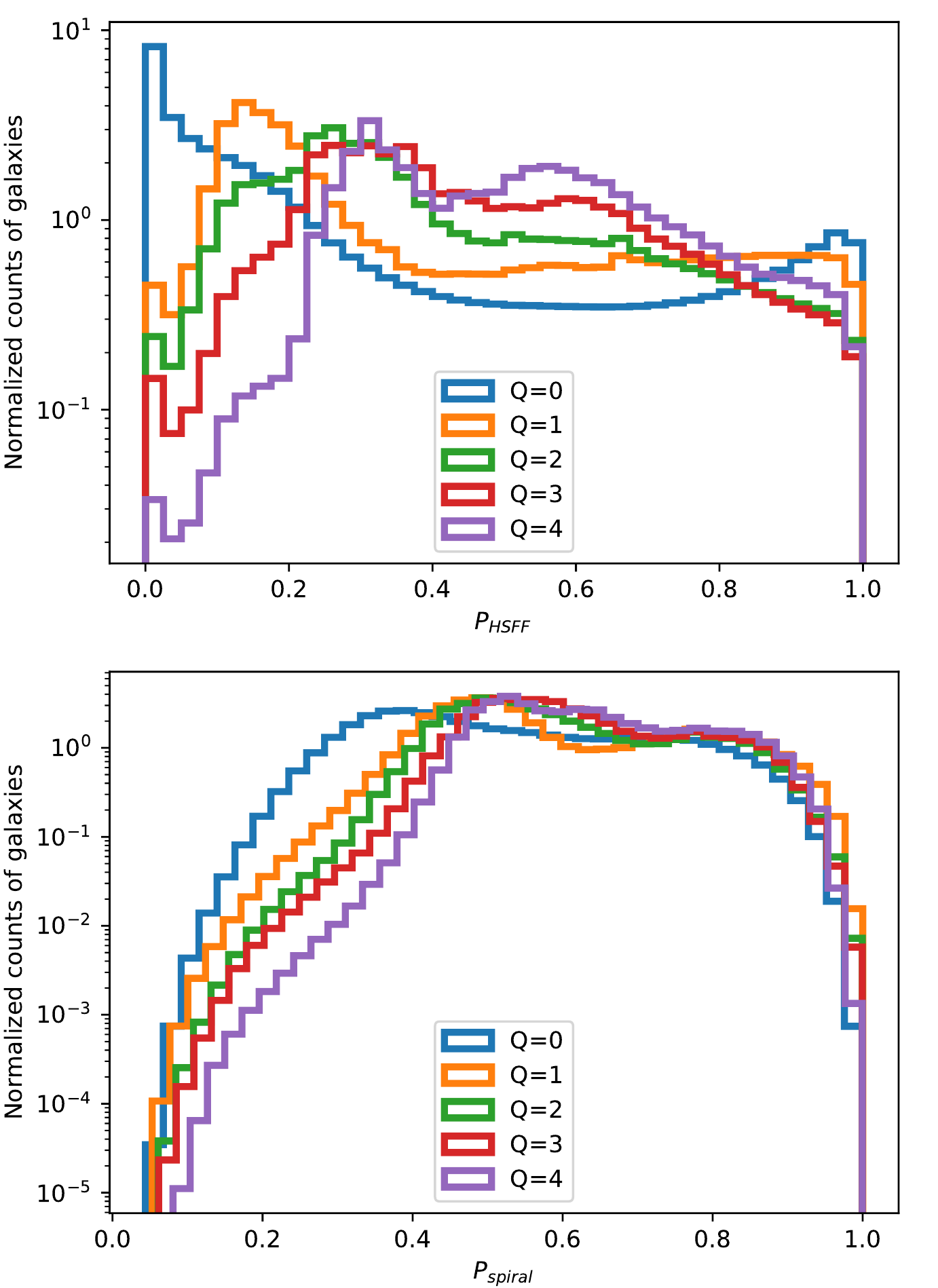}
%\vskip -1.5 cm
\caption{Distribution of the classification probability for PS1-DR2 galaxies (i.e. sources with $P_{\star}<0.2$) of having HSFF ($P_\mathrm{HSFF}$, upper panel) and of being spiral ($P_\mathrm{spiral}$, lower panel) for different values of $Q$. We estimate the classification probability with the RF model developed in \S\ref{RFNYU}  and \S\ref{preprocess} for the upper and lower panel, respectively. Large $P_\mathrm{HSFF}$  suggests a  high probability that the galaxy is HSFF.  Similarly, large $P_\mathrm{spiral}$ suggests a high probability that a galaxy is S. Larger font for axis labels. Poorer data sets ($Q>0$) tend to be associated with  probability values closer to random guessing $P=0.5$, as expected.} 
\label{fig:100}
\end{figure}

Next we  discuss the issue of missing data in PS1-DR2. For each source in the catalog of Appendix \ref{catalog_desc} we add a data-quality flag $Q$, where $Q=0$ indicates a  data set with complete information for all the five Pan-STARRS photometric filters, while $Q>0$ indicates that some data are missing. PS1-DR2 offers two flux measurements (i.e. PSF and Kron) for each of the five filters, for a total of 10 flux measurements per source with complete data. In the following, the value of the $Q$ variable quantifies the number of missing flux measurements in any filter (so that, for example, $Q=1$ means that one flux measurement is missing, etc.).  More details on the quality flag column are provided in Appendix \ref{catalog_desc}. For sources with  $Q>0$ we fill in the missing information by employing a linear interpolation of the spectral energy distribution. $53 \%$ of  PS1-DR2 sources  have  $Q=0$. The classification probability of the RF model for objects with $Q>0$ should be treated with caution. We quantify this statement below.

Figure \ref{fig:100} (upper panel) shows the distribution of the RF classification probability for PS1-DR2 galaxies to have HSSF. 
 PS1-DR2 galaxies with complete information (i.e. $Q=0$) are characterized by a bimodal classification probability distribution with one peak around $0$ and a second peak around $1$. This result suggests that galaxies with complete data are reliably classified as either having HSFF or LMSFF with high confidence. %having a high confidence of being either  HSFF or LMSFF. 
Instead, galaxies with $Q>0$ are more clustered around the region of random guessing $P_\mathrm{HSFF}=0.5$, as expected from their poorer data quality.  The median  classification confidence of PS1-DR2 galaxies with HSFF or LMSFF is as follows: galaxies  with $Q=0$ ($Q>0$) are classified with median classification confidence of 0.9 (0.75). Specifically, galaxies  with $Q=1,2,3,4$ are classified with decreasing median classification confidence of 0.81, 0.72, 0.67, 0.65, respectively. We note that the larger median classification confidence in the cross-validation set of 0.94 of \S\ref{RFNYU}  originates from the fact that the cross-validation set is biased towards brighter sources, which are easier to classify. Another key difference between the two sets is that the sources in the cross-validation set are all galaxies, while the application set has some level of contamination by stars
even after we filter on $P_{\star}$. A more detailed description of this effect is provided in \ref{extinction}.

In Figure \ref{fig:100}, lower panel, we perform a similar exercise for $P_\mathrm{spiral}$ and we compute the classification probability for different $Q$ values. As before, data sets with missing values are associated with classification probabilities more clustered around the value of random guessing, as expected (median of 0.64 for $Q=0$, and median of 0.59 for $Q>0$). $P_\mathrm{spiral}$  never approaches 0. The minimum value of $P_\mathrm{spiral}$ in the sample specify which sample is 0.04. This is likely due to the difference between the cross-validation set and the whole PS1-DR2 data set.

\begin{figure}
\centering
\includegraphics[width=0.5\textwidth]{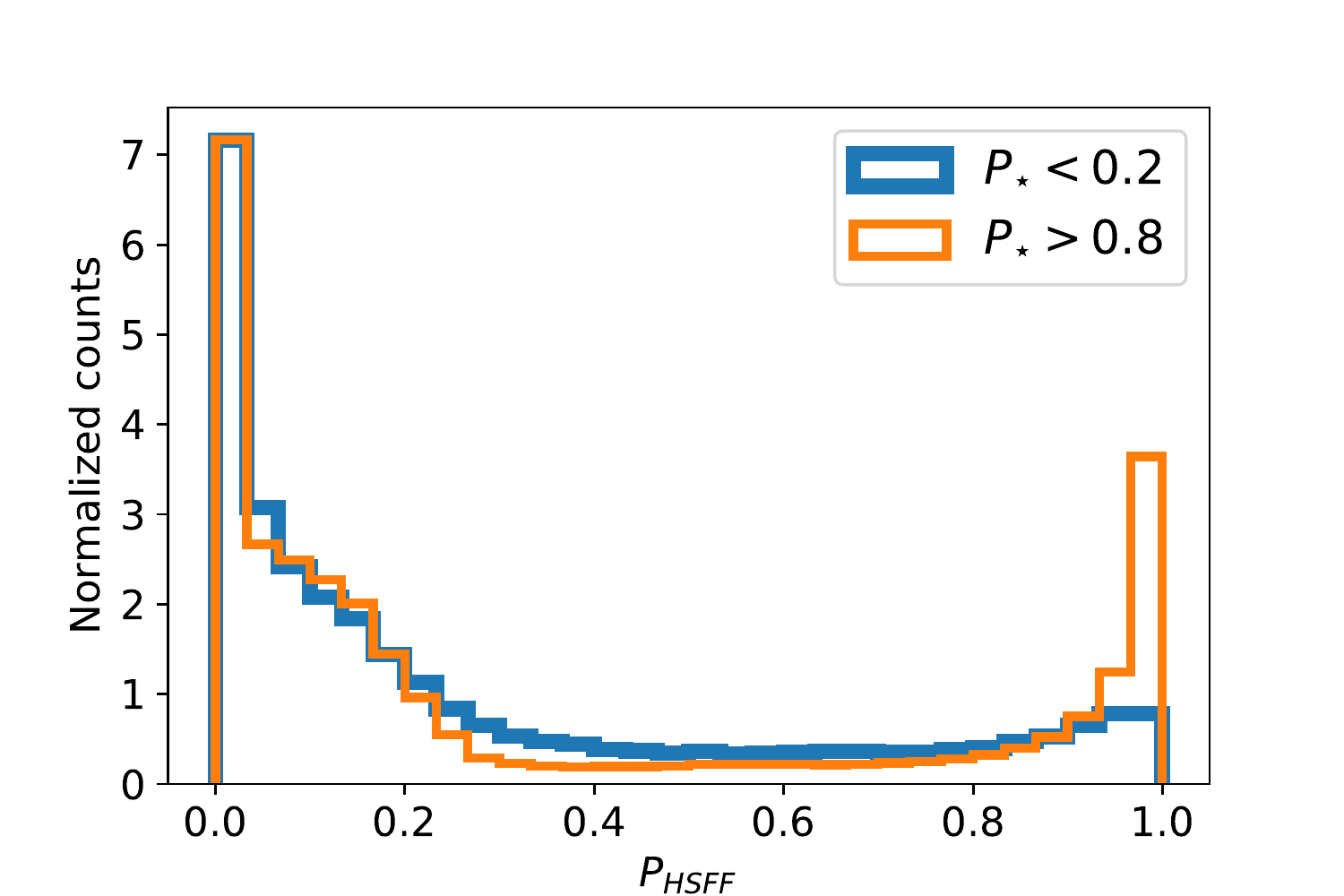}
\caption{ Classification probability distribution 
of PS1-DR2 sources of having HSFF for different cuts on $P_{\star}$. Blue and orange lines are used for   $P_{\star}<0.2$  and $P_{\star}>0.8 $, respectively. Note that the two distributions are slightly different at  $P_\mathrm{HSFF}>0.9$. }
\label{fig:121}
\end{figure}

Finally, in Figure \ref{fig:121} we explore the effect of increasing the sample contamination with stars by applying different cuts on $P_{\star}$.   
We consider two different scenarios: sources with $P_{\star}<0.2$ (i.e., most likely galaxies) and sources with $P_{\star}>0.8$ (i.e., most likely stars).  The  classification probability distributions of Figure \ref{fig:121} show some level of dependency on the cut on $P_{\star}$.
Not surprisingly, there are Galactic stars with colors able to perfectly mimic the colors of both highly star forming and quiescent galaxies.
For sources that are likely Galactic stars (i.e. with large $P_{\star}$) the $P_\mathrm{HSFF}$ value is also very likely to be meaningless. Taken at face value, sources with $P_{\star}>0.8$ are more likely to be considered to have HSFF than sources with $P_{\star}<0.2$. Therefore, we recommend to associate a physical meaning to $P_\mathrm{HSFF}$ and $P_\mathrm{spiral}$ only for sources with small values of $P_{\star}<0.2$ (outside the plane of the Galaxy).
Therefore, we recommend only using  the $P_\mathrm{HSFF}$ and $P_\mathrm{spiral}$ values of sources that have small values of $P_{\star}<0.2$ (outside the plane of the Galaxy)\footnote{More detailed user guidelines are provided in Appendix \ref{tab1}.}
%that only the sources with small values of $P_{\star}$ should be considered, irrespective of the nominal value of $P_\mathrm{HSFF}$. 
We conclude by noting that despite intrinsic differences between the training/testing data set and the entire PS1-DR2 catalog our algorithm is able to achieve a large median classification confidence of $\sim$0.9 for galaxies  with complete data in the PS1-DR2 release (i.e. with $Q=0$).

%-------------------------------------------------------------------
\subsection{Effects of Galactic Extinction}
\label{extinction}

Galactic extinction impacts the classification confidence of PS1-DR2 sources, as it directly affects the observed colors of celestial objects outside the Galaxy. In this section we discuss the effects of Galactic extinction on our classification capabilities and we quantitatively explore the possibility of applying an extinction correction to PS1-DR2 data to improve on the performance of our algorithm at low Galactic latitudes. Galactic extinction has no effect on the training/testing sets because the data contains only  objects  at high Galactic latitudes $|b|> 15^\circ$, which are  not significantly affected by Galactic reddening. In this section we explore the effects of Galactic reddening on our classification performance using the HSFF vs.\ LMSFF classifier of \S\ref{RFNYU}. Analogous results hold for the morphology classifier of \S\ref{preprocess}. For this reason, we do not include a specific discussion on the effects of the Galactic extinction correction for the RF morphology classifier (which is also outperformed by the HSFF vs.\ LMSFF classifier, as discussed below). 

\begin{figure}
\centering
\includegraphics[width=0.52\textwidth]{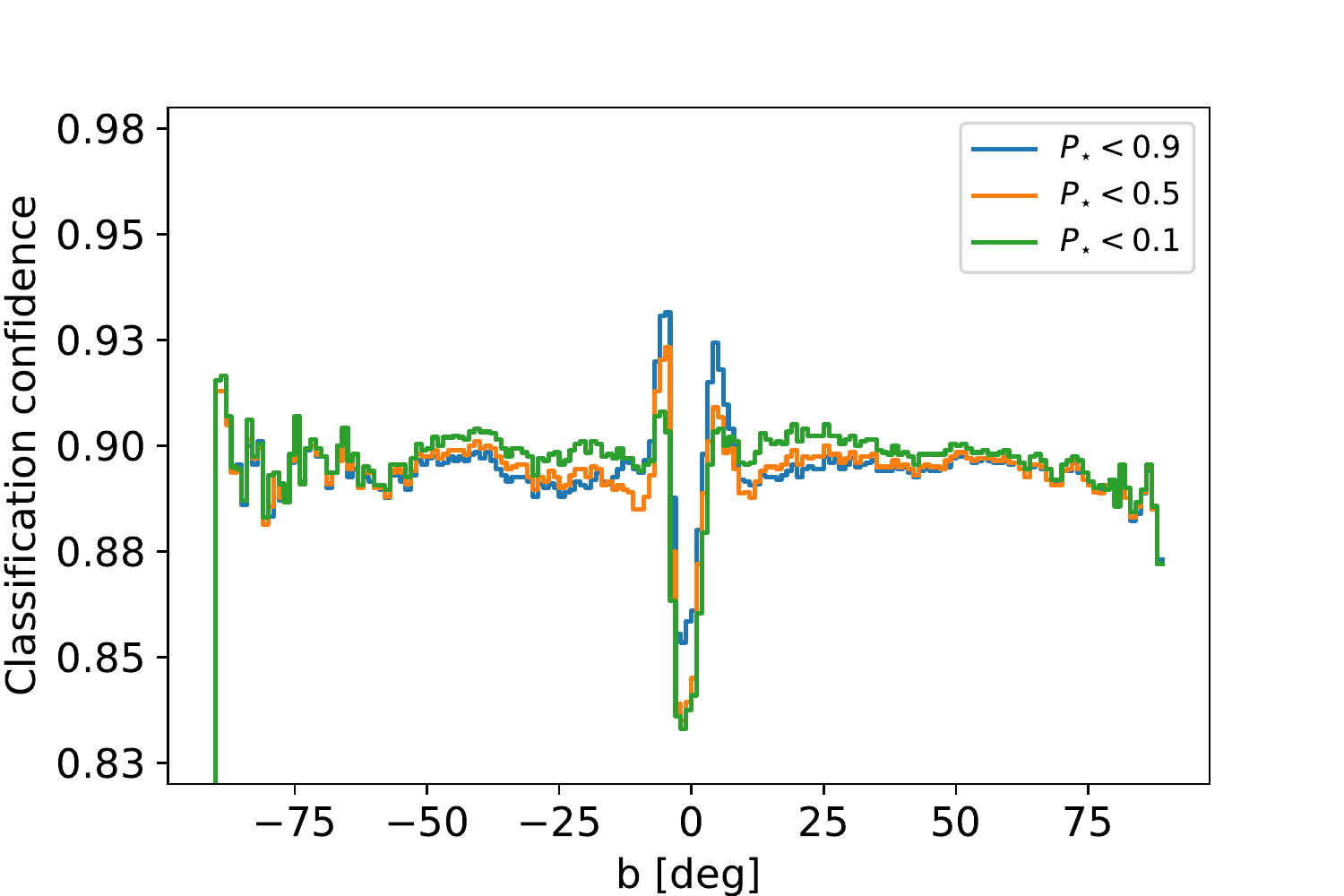}
\includegraphics[width=0.52\textwidth]{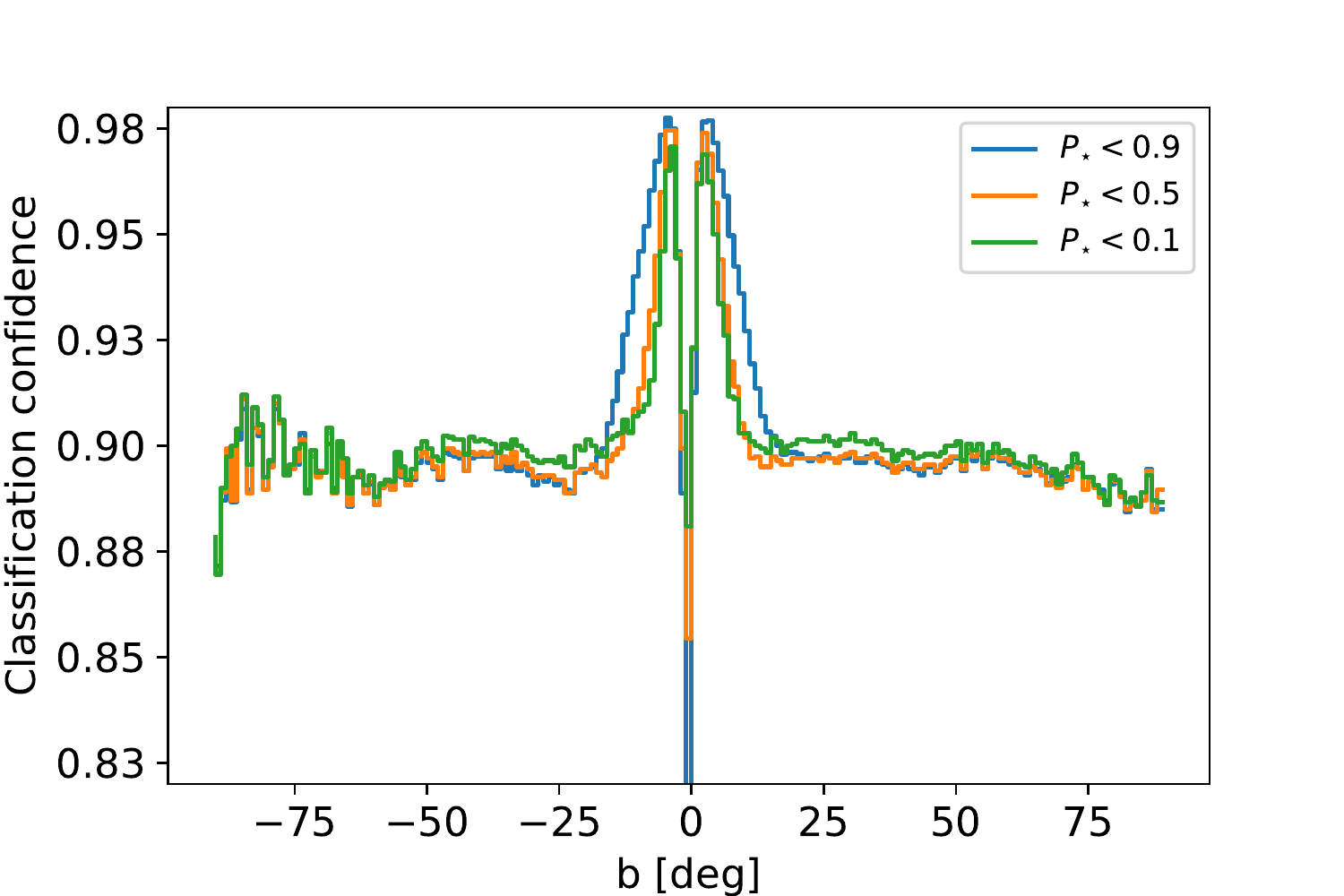}
\caption{Classification confidence for the PS1-DR2 sources of having HSFF before (upper panel) and after (lower panel) Galactic extinction correction. We plot the classification confidence for different values of $P_{\star}$  ($P_{\star}<0.1$, $P_{\star}<0.5$, $P_{\star}<0.9$) and for $Q=0$. \emph{Upper panel:} for low Galactic latitudes $|b|\lesssim 8^{\circ}$ the classification confidence differs significantly from  the mean value due to the large Galactic reddening. By constraining on the non-stellar nature of the objects of interest $P_{\star}<0.1$  results in less pronounced peaks, and vice versa.
 \emph{Lower panel:} applying the Galactic extinction correction has the effect of mitigating the large decrease of classification confidence around $b\sim0^{\circ}$. However, it also produces more pronounced   ``wings'', which are likely associated with contamination by stellar objects.}
\label{fig:130}
\end{figure}

We explore the performance of our classification algorithm as a function of Galactic latitude $b$ in Figure \ref{fig:130}, where we plot the classification confidence 
of PS1-DR2 sources (for $P_{\star}<k$ with $k=0.1,0.5,0.9$) with complete data. We find that before applying any extinction correction for low Galactic latitudes in the range $-8^{\circ} \lesssim b\lesssim 8^{\circ}$  the classification confidence significantly differs from the mean value, and lies below the mean value at $|b|\lesssim 1^{\circ}$, reaching a minimum of $\sim0.84$ (Figure \ref{fig:130}, upper panel). 
Furthermore, there are two peaks of high classification confidence around $b= -5.5^{\circ}$ and $b=5.5^{\circ}$ that result from the combined effects of high extinction and large contamination by stars. The amplitude of these two peaks is sensitive to the assumed upper cut on  $P_{\star}$ (and hence to the allowed level of contamination by stellar objects).
By constraining on the non-stellar nature of the objects of interest $P_{\star}<0.1$ directly results in less pronounced peaks, and vice versa (Figure \ref{fig:130}, upper panel).

One  way of looking at this result is that in the absence of any extinction correction, along the Galactic plane our algorithm very confidently mistakes highly reddened stars as early-type galaxies. Therefore, we may infer that the peaks originate from the fact that there is a \emph{certain amount} of extinction that makes reddened stars mimic the colors of early-type galaxies. This effect happens at a small range of Galactic latitudes.  At even lower Galactic latitudes the reddening is more extreme, and the reddened stars no longer look like early-type galaxies. This explains the location and presence of the peaks, the presence of the deep minimum at very small Galactic latitudes, and the fact that by filtering out stars we remove the peaks. In this latter case we are changing the underlying  colors of the population, that  is no longer a population of stars, but galaxies, which implies that that amount of galactic reddening will no longer be able to accurately mimic the colors of an early-type galaxy.

\begin{figure}
%\centering
\includegraphics[width=0.55\textwidth]{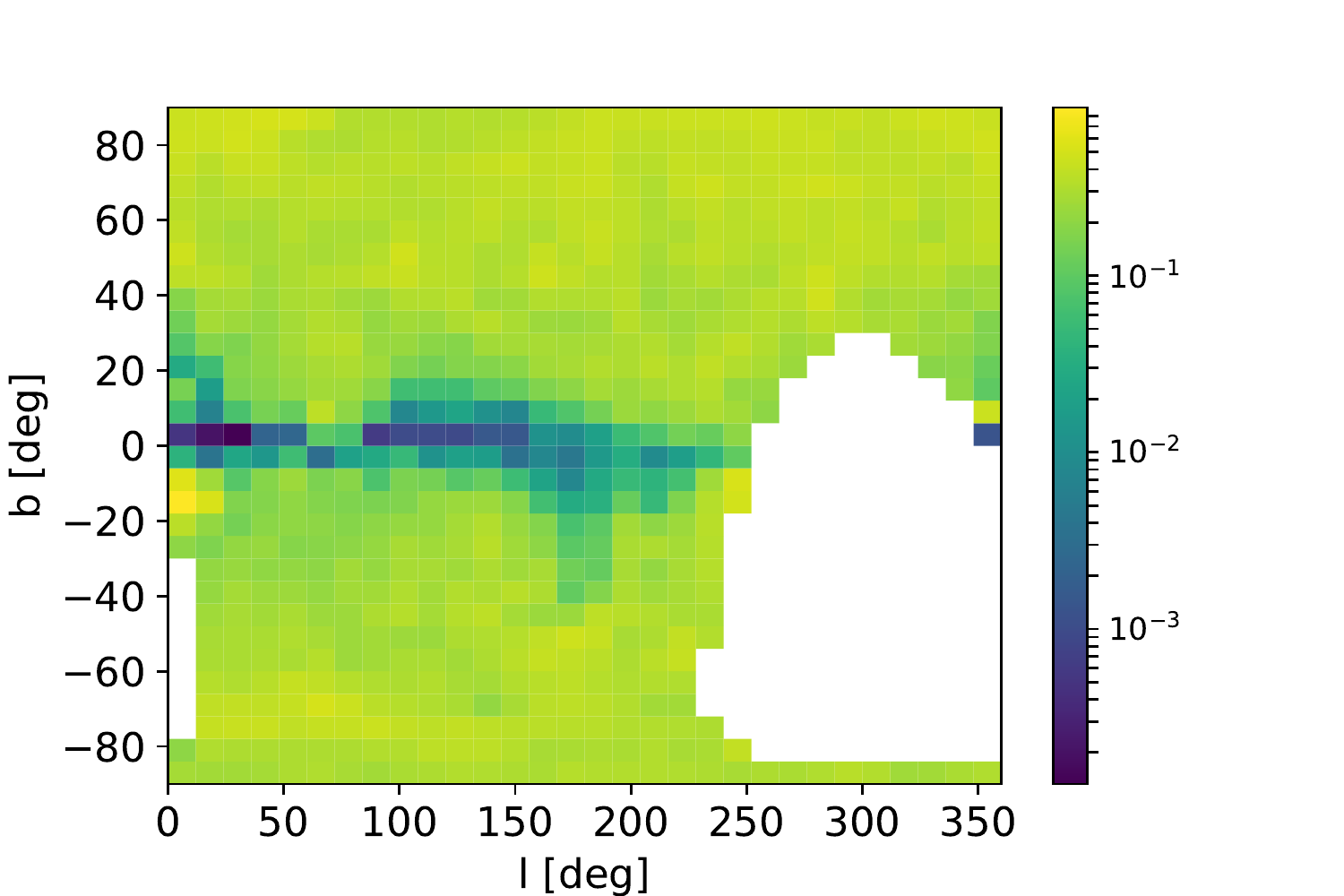}
\caption{Bi-dimensional histogram of the ratio between the
number of  HSFF galaxies (here defined as $P_\mathrm{HSFF}>0.6$) and  LMSFF galaxies ($P_\mathrm{HSFF}<0.4$) for PS1-DR2 galaxies ($P_{\star}<0.2$)  as a function of galactic coordinates. The white region is outside the PS1-DR2 footprint (corresponding to $\delta<-30^{\circ}$).  The $HSFF/LMSFF$ ratio in the Galactic plane is smaller than in the rest of the sky. This effect is an expected result from severe Galactic reddening, which artificially reduces the number of objects with observed blue colors. 
}
\label{fig:13}
\end{figure}

We further visualize the impact of Galactic extinction on our classification in Figure \ref{fig:13} by showing the bidimensional distribution of the ratio between the number of HSFF galaxies (defined here as $P_\mathrm{HSFF}>0.6$) and galaxies with LMSFF ($P_\mathrm{HSFF}<0.4$) in the sky. As before, for this test we select galaxies ($P_{\star}<0.2$) with complete information ($Q=0$).  
As expected, in proximity to the Galactic plane the $HSFF/LMSFF$ ratio is significantly lower than average. In this region of the sky the number of sources classified as LMSFF is significantly (and artificially) larger due to the redder observed colors.

Next, we quantify the amount of Galactic extinction along the line of sight for each source in the  PS1-DR2 using the extinction map by \citet[][]{Schlafly2011}. We then run our RF classification algorithm on the extinction corrected PS1-DR2 photometry and we compare our galaxy-classification results to the pre-extinction correction results. We are also aware that most of the sources in the Galactic plane are stars for which the extinction correction along the line of sight is only approximate.
Figure \ref{fig:130}, lower panel, shows the resulting classification confidence after the Galactic extinction correction has been applied. We find that while the depth and width of the absolute minimum of the classification confidence at $|b|\sim0^{\circ}$ seem to benefit from the extinction correction, the two peaks at $|b|\sim5^{\circ}$ are largely unaffected (if not even strengthened). The larger width of the peaks at $|b|\sim5^{\circ}$ most likely results from the fact that we are artificially creating more stars with the same colors as early-type galaxies.  We conclude that the anomalous behavior of the classification confidence around the Galactic plane is mainly driven by a large contamination of stars. Since we do not obtain significantly better performance with extinction corrections, we present the classification catalog without applying any Galactic extinction correction and we advise the user to be very selective on $P_{\star}$ especially  in the Galactic plane. A reasonable cut may be $P_{\star}<0.2$ outside the Galactic plane ($b<-8^\circ$ or $b>8^\circ$) and $P_{\star}<0.1$  in the plane  (here defined as $-8^\circ  <b< 8^\circ$). We also suggest trusting more our inferences on sources  with $Q=0$.

\begin{figure*}
\centering
\includegraphics[width=0.86\textwidth]{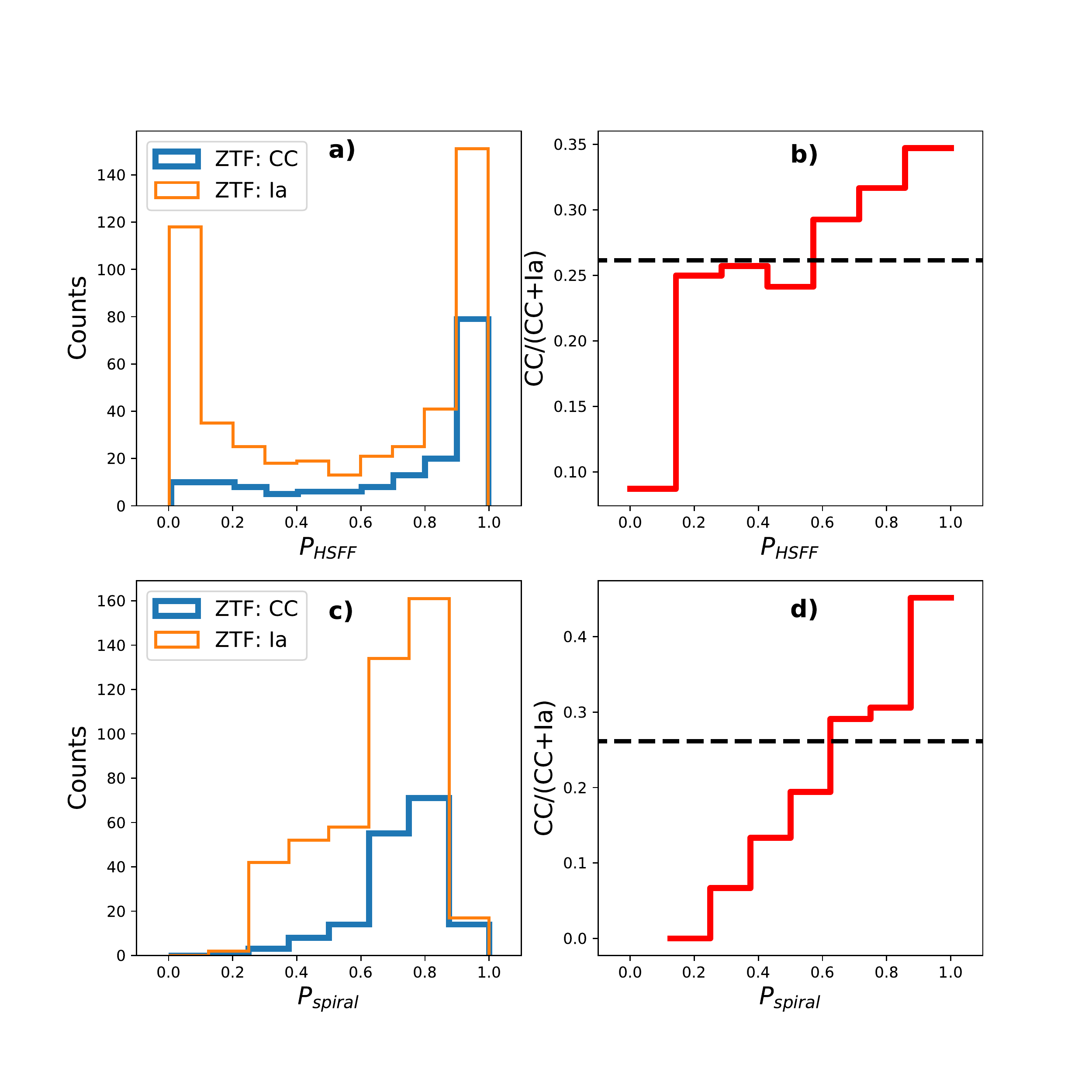}
\vskip -40pt
\caption{Panel (a):  $P_\mathrm{HSFF}$ distribution  of the  host galaxies of SNe from the ZTF-BTS. Orange (Blue) thick lines show galaxies associated with type Ia SNe (CC-SNe). Panel (b) red line:  fraction  of CC-SNe ($CC/(CC+Ia)$) as a function of $P_\mathrm{HSFF}$. Horizontal black dashed line: fraction of CC-SNe in the ZTF sample that we consider here, which corresponds to the random guessing level. Panel (c): $P_\mathrm{spiral}$ distribution of SN host galaxies from the ZTF magnitude-limited catalog.  Panel (d), red line:  fraction of CC supernovae ($CC/(CC+Ia)$) as a function of $P_\mathrm{spiral}$. In panels (b) and (d) any significant departure of the $CC/(CC+Ia)$ fraction from the horizontal black dashed line can be considered an improvement over random guessing. As discussed in \S\ref{panst} the $P_\mathrm{spiral}$ score never reaches 0 because of the intrinsic difference between the cross-validation and the entire PS1-DR2 data set.}
\label{fig:ration}
\end{figure*}

\begin{figure*}
\centering
\includegraphics[width=0.86\textwidth]{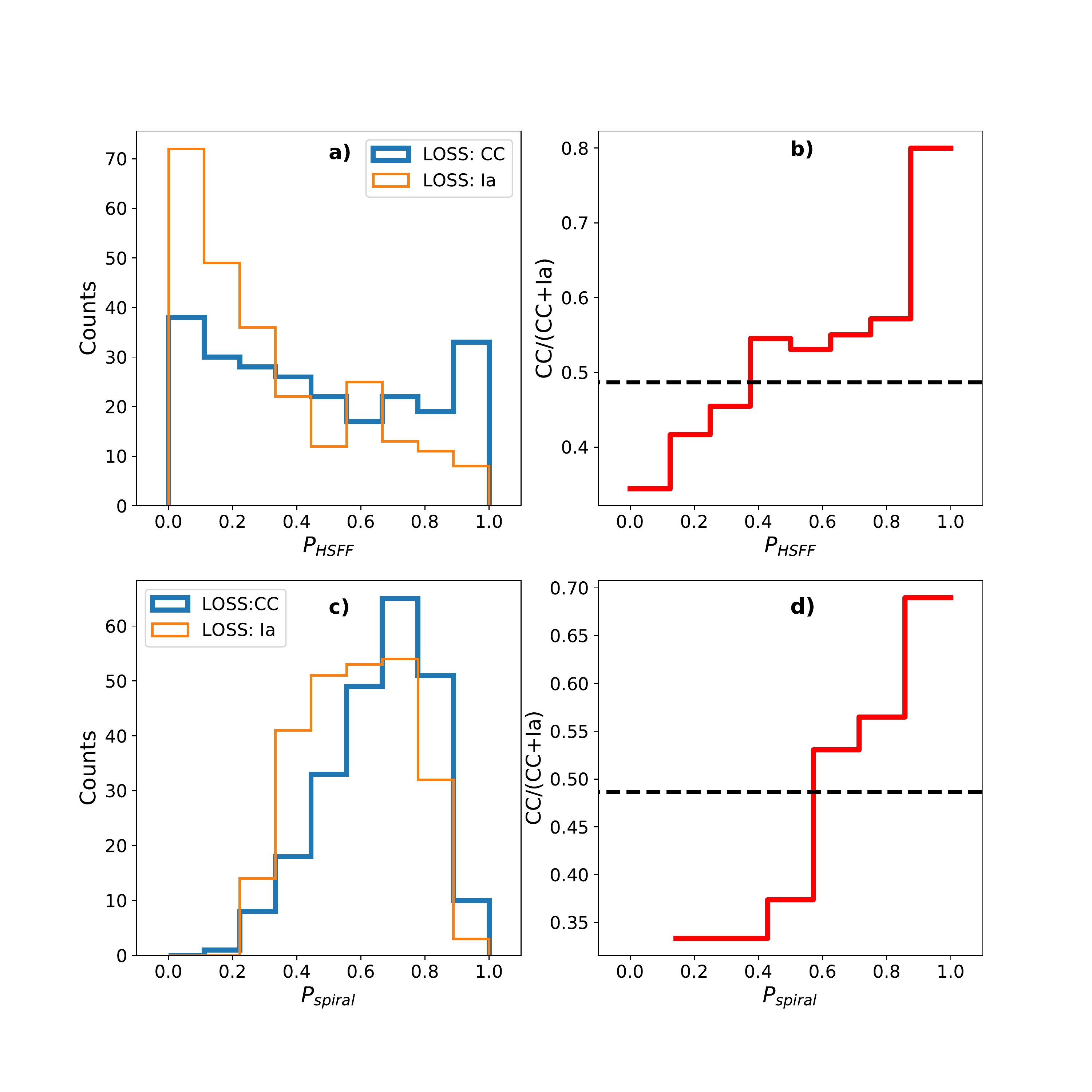}
\vskip -40pt
\caption{Panel (a):  $P_\mathrm{HSFF}$ distribution of host galaxies  of SNe from the  galaxy-targeted LOSS. Orange (Blue) thick lines: galaxies associated with type Ia SNe (CC-SNe). Panel (b), red line:  fraction  of CC-SNe ($CC/(CC+Ia)$) as a function of $P_\mathrm{HSFF}$. Horizontal black dashed line: fraction of CC-SNe in the LOSS sample that we study here, which corresponds to the random guessing level. Panel (c): $P_\mathrm{spiral}$ distribution of host galaxies of SNe from the LOSS  catalog.  Panel (d), red line:  fraction  of CC supernovae ($CC/(CC+Ia)$) as a function of $P_\mathrm{spiral}$. In panels (b) and (d) any significant departure of the $CC/(CC+Ia)$ fraction  from the horizontal black dashed line may be considered an improvement over random guessing. 
We interpret the ``flat'' distribution of $P_\mathrm{HSFF}$ of CCSN host-galaxies as due to the fact that the galaxies in the training sample have significantly larger distances than  the nearby galaxies targeted by the LOSS.
}
\label{fig:ration2}
\end{figure*}
%--------------------------------------------------------------------------------------------
%\subsection{Classification of P\lowercase{an}-STARRS sources based on their morphology}
%\label{morpho}

%%%%%%%%%%%%%%%%%%%%%%%%%%%%%%%%%%%%%%%%%%%%%%%%%%%%%%%%%%%%%%%%%%%%
\section{Using  $P_\mathrm{HSFF}$ and $P_\mathrm{spiral}$  for prompt supernova classification}
\label{transient}
The immediate goal of this paper is to characterize the star formation properties of galaxies within  PS1-DR2. In this section we carry out a simple exercise that serves as a proof of concept to highlight the predictive power of $P_\mathrm{HSFF}$ and $P_\mathrm{spiral}$ in the 
context of supernova (SN) typing (core-collapse vs thermonuclear SNe) at the time of their first detection. This is part of a larger effort aimed at classifying transients by combining information on the transient's environment and their photometric evolution.
Specifically, we will show how the $P_\mathrm{HSFF}$ and $P_\mathrm{spiral}$  scores can be used to statistically infer the SN type. As these scores are available from pre-explosion PS1 images and can be readily associated to the host-galaxy of a newly identified transient,  these scores may be used as a useful tool to improve our capabilities of  prompt classification of transients at the time of their first detection. A detailed analysis of the host-galaxy properties and their connection to the transient properties is beyond the scope of this work and will be addressed in a forthcoming paper. Here we focus our analysis on the relation between the star formation properties and morphological properties of a  galaxy and the probability that it will host core-collapse or thermonuclear stellar explosions. We will use SN spectroscopic classifications from both a magnitude-limited untargeted transient survey (Zwicky Transient Facility, ZTF), and a galaxy-targeted nearby supernova survey (Lick Observatory Supernova Search, LOSS). 

We start by considering spectroscopically classified SNe from the  ZTF Bright Transient Survey (BTS, \citealt{Fremling2019}). The ZTF-BTS contains  transients brighter than 18.5 mag at peak, at a distance corresponding to  $z \lesssim 0.15$ ($d\lesssim 700$ Mpc). 
We associate each spectroscopically classified SN in the ZTF-BTS catalog with its host galaxy  in PS1-DR2 ( and its respective $P_\mathrm{HSFF}$ and $P_\mathrm{spiral}$ scores). We carry out the host/SN association with a  method developed by Stroh et al, in prep., which is based on \citet{Bloom2001}. The  ZTF-BTS catalog provides the PS1-DR2 host galaxies. Here we  use the Stroh et al. association method as we plan to expand this work to other transients and other surveys. This completely automatic procedure leads to a an association that is consistent with the host galaxies provided by ZTF-BTS. This algorithm identifies the likely host galaxy as the galaxy with the lowest chance coincidence probability ($P_{cc} = 1 - e^{\pi R_e^2 \sigma(\le m)})$ where $\sigma(\le m)$ is the galaxy number density as given by \citet{Berger2010}.
Following \citet{Blanchard2016}, the effective radius, $R_e=\sqrt{R^2 +(2.5R_{kron})^2}$ where $R$ is the angular separation between the ZTF host galaxy position and the PS1 DR2 potential host galaxy, while $R_{kron}$ is the PS1 DR2 g Kron radius of the galaxy.
We selected host galaxies with complete data in PS1-DR2 and obtain a sample of 162 core-collapse SNe (CCSNe, including types II, II-87A, IIb, IIn, Ib, Ib/c, Ibn, Ic, Ic-BL, Ic-pec, SLSN-I, SLSN-II)  and 464 thermonuclear SNe ``Ia'' in short, including branch-normal Ia, Ia-02cx, Ia-91T, Ia-91bg, Ia-SC, Ia-CSM). The final sample contains $26\%$ CCSNe by number. The median SN distance of the sample is $\sim$250 Mpc.

Figure \ref{fig:ration} (panel \emph{a}) shows the distribution of  $P_\mathrm{HSFF}$  for the host galaxies of CCSNe and type Ia SNe. Most host galaxies of type Ia SNe have $P_\mathrm{HSFF}<0.1$ or $P_\mathrm{HSFF}>0.9$, 
while CCSNe are mainly associated with actively star forming galaxies with $P_\mathrm{HSFF}>0.9$. In Figure \ref{fig:ration}, panel \emph{b}, we plot the fractional number of CCSNe (i.e. the ratio between the number of CCSNe and the total number of SNe)
as a function of $P_\mathrm{HSFF}$. As CCSNe constitute $\sim26\%$ of the sample, this fraction indicates the level one would obtain by random guessing (indicated by the  horizontal dashed line in Figure \ref{fig:ration}). This figure shows that if we select galaxies with $P_\mathrm{HSFF}<0.1$, the fraction of CCSNe drops to $CC/(CC+Ia)\sim7\%$. At higher $P_\mathrm{HSFF}$, the fraction of CCSNe increases. In particular we find that galaxies with $P_\mathrm{HSFF}>0.8$ have a large CCSNe fraction of $CC/(CC+Ia)\sim30-35\%$.
In  Figure \ref{fig:ration}  (panels \emph{c} and \emph{d}) we perform a similar exercise using $P_\mathrm{spiral}$.
We find that   the host galaxies of Ia SNe cover a wide range of $P_\mathrm{spiral}$ (i.e. type Ia SNe are hosted in early and late-type galaxies), while CCSNe are mainly hosted in galaxies with large $P_\mathrm{spiral}$, as expected. 
As before, the $CC/(CC+Ia)$ fraction is a monotonically increasing function of the $P_\mathrm{spiral}$ score. Specifically, for $P_\mathrm{spiral}>0.9$ we find a ratio of $CC/(CC+Ia)\approx 50\%$, while for $P_\mathrm{spiral}<0.4$ the ratio is $CC/(CC+Ia)\approx 10\%$.  

We perform a similar analysis using the nearby SN sample from LOSS, which is a galaxy-targeted search for SNe in the local Universe at $d\le 200$ Mpc \citep{Leaman2011}. This sample includes SNe that are significantly closer than those in the ZTF sample.  As before, we associate each SN with its host galaxy in PS1-DR2, and the relative $P_\mathrm{HSFF}$ and $P_\mathrm{spiral}$ scores.
The original LOSS sample of spectroscopically classified SNe consists of 929 sources. Of these, we selected 517 associations with good-quality PS1-DR2 photometry ($Q=0$). Our final sample consists of 249 type Ia SNe and 268 CCSNe (i.e. the CCSN fraction by number is $\approx 48\%$).

Figure \ref{fig:ration2} (panels \emph{a} and \emph{c}) shows the $P_\mathrm{HSFF}$ and $P_\mathrm{spiral}$ distributions of the host galaxies in our sample, while panels \emph{b} and \emph{d} show the fractional number of CCSNe ($CC/(CC+Ia)$) as a function of both scores.
As for the ZTF sample, $CC/(CC+Ia)$ is a (mostly) monotonic function of $P_\mathrm{HSFF}$ and $P_\mathrm{spiral}$.  We interpret the roughly ``flat'' distribution of $P_\mathrm{HSFF}$ of CCSN host-galaxies as due to the fact that the galaxies in the training sample have significantly larger distances than  the nearby galaxies targeted by the LOSS. Indeed, the most distant SN in the LOSS sample (at $d\sim200$ Mpc) is closer than $\sim90\%$ of the galaxies in the training set. 
As a consequence, in the training sample the relation between the Kron and PSF photometry is different to that of the large, well-resolved, nearby galaxies in LOSS. Any significant difference between the training set and the actual sample leads the algorithm to ``confusion'', the manifestation of which is this flat distribution of scores. With this caveat in mind, it is still interesting to note that $CC/(CC+Ia)\sim80\%$ for $P_\mathrm{HSFF}>0.9$ (which is significantly above the $52\%$ value expected for random guessing), and that the fraction of CCSNe is suppressed to $CC/(CC+Ia)\sim40\%$ for host-galaxies with $P_\mathrm{HSFF}<0.2$.  As expected, the $P_\mathrm{spiral}$ distribution is skewed towards large values for CCSNe.  For $P_\mathrm{spiral}<0.4$ we find $CC/(CC+Ia)\sim 33\% $ below the $52\%$ value of random guessing), while at large $P_\mathrm{spiral}>0.85$ the fraction of CCSNe is highly enhanced to $CC/(CC+Ia)\sim70\%$.

The predictive power of the host galaxy morphology in SN typing (CC vs. Ia SNe) was first quantified by \cite{Foley+2013} on the LOSS sample. \citet{Foley+2013} (their Figure 1, upper panel) showed that in their ``full'' LOSS sample which contains 41\% of Ia SNe by number, the fraction of SNe Ia in E/S0 galaxies is in the range $\sim$65-100\%  (corresponding to $CC/(CC+Ia)\sim 0-35\%$), decreasing to $\lesssim20$\% (or $CC/(CC+Ia)\gtrsim 80\%$) in Sbc/Sb/Scd/Irr galaxies. While it is not possible to directly compare our results to  Figure 1 of \citet{Foley+2013}, it is interesting to note that our RF classifiers that are uniquely based on host-galaxy colors reach comparable purity levels at the extremes of the $P_{\rm{spiral}}$ or $P_{\rm{HSFF}}$ distributions. 
\citet{Foley+2013} further employed a Naive Bayes classifier that leverages the transient's contextual information such as the host galaxy morphology, absolute magnitude ($M_k$), colors ($B_0-K$), offset from host-galaxy nucleus and pixel rank for type Ia SN identification.
Their Naive Bayes classifier returns the probability ($p_{Ia}$) for each LOSS SN of being a Ia SN. These authors found that 30$\%$ of SNe in their sample have $p_{Ia}>0.5$; of these, $71\%$ are SNe Ia. This result compares favorably to the random guessing level  of $P(Ia) = 41\%$. For the same sample, $21\%$ of SNe have $p_{Ia}<0.1$, $84\%$ of which are CC SNe. These findings are the result of the combination of inferences obtained from the different sources of contextual information listed above (including detailed host-galaxy morphology classification, the transient's distance, and absolute magnitudes $M_k$). These features are not  available in main wide field transient surveys, implying that the \citet{Foley+2013} methodology in its current form can not be easily extended to the very large data sets like those of the LSST. Simplified approaches that rely on minimal contextual information (e.g., colors) have the advantage that they are directly applicable to most transients surveys.

The important conclusion from these two exercises on the LOSS and ZTF-BTS samples is that by selecting on the $P_\mathrm{HSFF}$ or $P_\mathrm{spiral}$ scores of SN host galaxies it is possible to artificially and significantly enhance or suppress the fraction of CCSNe (or thermonuclear SNe) with respect to random guessing. This result demonstrates that it is possible to improve on the SN classification at the time of their first detection by using the available information on their large-scale environments processed with machine learning algorithms (i.e. no-human in the loop).

\section{Summary and Conclusions}
\label{conclusions}

Machine learning is becoming a fundamental tool in a variety of fields in astrophysics, from exoplanet discovery to galaxy and transient classification. 
In this paper we have developed two machine learning algorithms and presented the classification of galaxies in the P\lowercase{an}-STARRS $3\pi$ survey based on their morphology and recent star formation history. Specifically, we have trained and tested two  random forest (RF) models on a sub-sample of the PS1-DR2 galaxies using PS1-DR2 colors as input features for the RF classifiers, and using labels from the Huertas-Company data-set (for galaxy morphologies) and from the New York University Value-Added Galaxy Catalog (NYU-VAGC, for the fraction of star formation occurred in the last 300 Myr). We have obtained a classification accuracy of $78\%$  when discriminating between elliptical and spiral galaxies in the cross validation set. The classification accuracy is $89\%$ when discriminating between galaxies with high and low-to-moderate star formation fraction (HSFF vs.\ LMSFF) in the cross validation set. We have then applied  both RF models  to the entire PS1-DR2 catalog to determine the probability that each galaxy is spiral ($P_\mathrm{spiral}$) and whether it has a HSFF ($P_\mathrm{HSFF}$) or not. We present our classifications in a catalog with a structure as outlined in Appendix \ref{catalog_desc}. User guidelines are also described in Appendix \ref{catalog_desc}.

We have  applied the two RF classifiers to host galaxies of two SN samples from the ZTF-BTS and LOSS, and we have demonstrated that the colors of the transient's host galaxies can be used to statistically infer their star formation  and morphological properties  in a way that can be used to aid transient classification at the time of the first detection (in line with the initial study by \citealt{Foley+2013}). The ZTF-BTS and the LOSS samples contain core-collapse SNe (CCSNe) and stellar explosion of thermonuclear origin. For both the ZTF-BTS \citep{Fremling2019} and LOSS \citep{Leaman2011} samples we find that $P_\mathrm{spiral}$ and $P_\mathrm{HSFF}$ are highly correlated with the fraction of CCSNe.

In particular, for the brightness-limited SN sample from ZTF-BTS, selecting host galaxies with $P_\mathrm{HSFF}>0.8$ we obtain a $\approx 10\%$ larger fraction of CCSNe with respect to random guessing, while for $P_\mathrm{HSFF}<0.1$ we obtain a $\approx 20\%$ lower CCSN fraction  with respect to random guessing. Furthermore, selecting host galaxies with $P_\mathrm{spiral}>0.9$ we obtain a $\approx 50\%$ fraction of CCSNe (which constitutes a $\sim24\%$ improvement with respect to random guessing). For the galaxy-targeted SN sample from LOSS we obtain similar results. In this case, $\sim70\%-80\%$ of SNe associated with likely spiral host galaxies ($P_\mathrm{spiral}>0.9$) or galaxies with high star formation fraction  ($P_\mathrm{HSFF}>0.9$) are of core-collapse origin, compared with the $48\%$ fraction of CCSNe in the sample.

Our work demonstrates that it is possible to achieve significant improvements in prompt SN classification by using available contextual information automatically processed with machine learning algorithms.  The host galaxy information from our catalog can thus be directly used to complement and improve the classification accuracy of existing algorithms that solely rely on the transient's photometric properties.  A key advantage of classifiers that will include inference from contextual information is related to the fact that (some of) the host galaxies properties are known at the time of the very first detection of a new transient, when the photometric information is exceedingly limited. In the current era of  spectroscopically-starved time-domain astronomy, the capability of promptly inferring the nature of a large number of transients without spectroscopic follow up (or visual inspection of each individual host galaxy) is of paramount importance. Indeed, in the near future, surveys like the Legacy Survey of Space and Time (LSST) carried out on the Vera C.  Rubin Observatory  will dramatically increase the discovery rate of transients by producing $\approx 10^6$ alerts per night, making a systematic transient spectroscopic-classification not viable. In a future paper, we will extend the use of contextual information for prompt transient classification to include other properties of the large-scale environments of a variety of astronomical transients.

\bigskip
\bigskip\bigskip\bigskip
\section*{Acknowledgments}
The authors thank Vicky Kalogera and Wen-fai Fong for discussions on this project.
This work is supported by the Heising-Simons Foundation under grant \#2018-0911 (PI: Margutti).
R.M. is grateful to KITP for hospitality during the completion of this paper. This research was supported in part by the National Science Foundation under Grant No. NSF PHY-1748958.
R.M.~acknowledges support by the National Science Foundation under Award No. AST-1909796. Raffaella Margutti is a CIFAR Azrieli Global Scholar in the Gravity \& the Extreme Universe Program, 2019. 
A.A.M.~is funded by the Large Synoptic Survey Telescope Corporation, the
Brinson Foundation, and the Moore Foundation in support of the LSSTC Data
Science Fellowship Program; he also receives support as a CIERA Fellow by the
CIERA Postdoctoral Fellowship Program (Center for Interdisciplinary
Exploration and Research in Astrophysics, Northwestern University).
The Pan-STARRS1 Surveys (PS1) and the PS1 public science archive have been made possible through contributions by the Institute for Astronomy, the University of Hawaii, the Pan-STARRS Project Office, the Max-Planck Society and its participating institutes, the Max Planck Institute for Astronomy, Heidelberg and the Max Planck Institute for Extraterrestrial Physics, Garching, The Johns Hopkins University, Durham University, the University of Edinburgh, the Queen's University Belfast, the Harvard-Smithsonian Center for Astrophysics, the Las Cumbres Observatory Global Telescope Network Incorporated, the National Central University of Taiwan, the Space Telescope Science Institute, the National Aeronautics and Space Administration under Grant No. NNX08AR22G issued through the Planetary Science Division of the NASA Science Mission Directorate, the National Science Foundation Grant No. AST-1238877, the University of Maryland, Eotvos Lorand University (ELTE), the Los Alamos National Laboratory, and the Gordon and Betty Moore Foundation.

\appendix
\section{Catalog of classifications of PS1-DR2 sources}
\label{catalog_desc}
The catalog is organized as follows. The first column is the ID of the P\lowercase{an}-STARRS object. The second and third columns are the RA and DEC coordinates measured in degrees. The fourth column represents the probability for an object to be a point-source ($P_{\star}$), as derived by \cite{Tachibana2018}. The fifth column represents the probability for a source to be HSFF ($P_\mathrm{HSFF}$). The sixth column represents the probability for a source to be spiral ($P_\mathrm{spiral}$).
Note that when $P_{\star}>0.5$ , the  values of $P_\mathrm{HSFF}$ and  $P_\mathrm{spiral}$ are meaningless.
The seventh column is a completeness flag for the data. PS1-DR2 offers ten flux density measurements for each source: five PSF fluxes (for the $g$, $r$, $i$, $z$ and $y$, respectively) and five Kron fluxes (one for each photometric band). The completeness flag is expressed as a ten digit binary number, where each digit tells if the data in a specific filter is present (0) or missing (1). The first five digits are related to the PSF fluxes for photometric bands in this order: $g$, $r$, $i$, $z$, $y$. The second five digits are associated with Kron fluxes for the same order of photometric bands. As reference, the binary number 0000000000 states that there are no missing data (parameter $Q=0$ in the paper), 0000010000 states that the $g$-band Kron flux is missing ($Q=1$), and 0100000010 states that the $r$-band PSF and the $i$-band Kron are missing ($Q=2$). Here $Q$ represents the number of missing filters for each sources.

We recommend to mostly trust $P_\mathrm{HSFF}$ and $P_\mathrm{spiral}$ classifications for sources with $Q=0$, $P_{\star}<0.2$ \emph{and} Galactic latitude outside the range $-8^\circ<b<8^\circ$.  Classifications of sources close to the Galactic plane and classifications of sources with $P_{\star}>0.2$ should be treated with caution. Users interested in supernova classification (CC vs thermonuclear)
with $P_\mathrm{HSFF}$ and $P_\mathrm{spiral}$ may use the results in Figure \ref{fig:ration} and \ref{fig:ration2} as guideline.  In Table \ref{tab1}, we report the first few rows of the catalog for display.

\begin{table}
\begin{center}
\begin{tabular}{|c|c|c|c|c|c|c|}
\hline
ID         &            RA   &    DEC &  $P_{\star}$   & $P_\mathrm{HSFF}$ &  $P_\mathrm{spiral}$ & Completness flag \\
\hline

115573500156866067 & 350.0157709 & 6.31298624 &  0.0167 & 0.3218 & 0.4726 & 0000010000 \\
115591651428609170 &  165.14281046 & 6.33225118 & 0.9958 & 0.0018 & 0.3720 & 0000000000 \\
115592747546664681 & 274.75467722 &  6.32846762 & 0.7686 & 0.6024 & 0.6924 & 0000011001 \\
115591851329443064 & 185.13275190 &  6.32728953 & 0.64258 & 0.8345 & 0.924 & 0000000011 \\
115582187787633432 & 218.77857803 & 6.31919698 & 0.01426 & 0.3189 & 0.587 & 0000000000 \\

\hline
\end{tabular}
\caption{Sample table of the released catalog. Description of the table in Appendix \ref{catalog_desc}} \label{tab1}
\end{center}
\end{table}
%\keywords{stars:neutron -- gravitational waves -- infrared: general}

%%%%%%%%%%%%%%%%%%%%%%%%%%%%%%%%%%%%%%%%%%%
\bibliographystyle{aasjournal}
\bibliography{master_sne}

%%%%%%%%%%%%%%%%%%%%%%%%%%%%%%%%%%%%%%%%%%%%%%%%%%%%%%%%%

\end{document}